\documentclass[onecolumn, aps, nofootinbib, prd, floats, floatfix, amsmath, amssymb, superscriptaddress, preprintnumbers, preprint]{revtex4-2} %

\usepackage{graphicx}  
\usepackage{dcolumn}   
\usepackage{chemarrow}
\usepackage{color}
\usepackage[dvipsnames]{xcolor}
\usepackage{mathrsfs}
\usepackage{float}
\usepackage{physics}
\usepackage{slashed}
\usepackage{latexsym}
\usepackage{comment}

\usepackage[colorlinks=true, linkcolor=blue, citecolor=purple]{hyperref}
\usepackage{ulem}
\usepackage{orcidlink}
\usepackage{appendix}

\usepackage{tabularx} 
\newcolumntype{C}[1]{>{\centering\let\newline\\\arraybackslash\hspace{0pt}}m{#1}}

\begin{document}

\title{Collider Spin Tomography with Missing Neutrinos}

\author{Tengyu Ai}
\email{tengyuai@pku.edu.cn}
 \affiliation{School of Physics, Peking University, Beijing 100871, China}

\author{Qi Bi}
\email{biqii@buaa.edu.cn}
\affiliation{School of Physics, Beihang University, Beijing 100191, China}

\author{Yuxin He}
\email{yuxinhe@pku.edu.cn}
\affiliation{School of Physics, Peking University, Beijing 100871, China}

\author{Zekun Li}
\email{lizekun@stu.pku.edu.cn}
\affiliation{School of Physics, Peking University, Beijing 100871, China}
	
\author{Jia Liu\orcidlink{0000-0001-7386-0253}}
\email{jialiu@pku.edu.cn}
\affiliation{School of Physics, Peking University, Beijing 100871, China}
\affiliation{State Key Laboratory of Nuclear Physics and Technology, Peking University, Beijing 100871, China}
\affiliation{Center for High Energy Physics, Peking University, Beijing 100871, China}

\author{Xiao-Ping Wang\orcidlink{0000-0002-2258-7741}}
\email{hcwangxiaoping@buaa.edu.cn}
\affiliation{School of Physics, Beihang University, Beijing 100191, China}

\preprint{CPTNP-2026-020}

\begin{abstract}
Missing neutrinos need not destroy collider spin tomography. We
formulate the visible measurement under kinematic ambiguities arising
from invisible particles as a coarse-grained positive-operator-valued
measure on the production spin density matrix. We show that information
loss is governed by the null space of the resulting visible-data map,
not by the number of kinematic solutions. 
In $e^+e^-\to\tau^+\tau^-\to\pi^+\pi^-+\nu\bar\nu$, the twofold ambiguity
leaves only the antisymmetric spin-correlation combination
$C_{nr}-C_{rn}$ unidentifiable, while the differential production rate
and the remaining fourteen spin coefficients are identifiable. For
practical reconstruction under kinematic ambiguities, we
develop a self-consistent fixed-point unfolding method using only
visible data, without assuming a theoretical production template.
Closure tests in Standard Model and anomalous tau-dipole benchmarks
show that the method reproduces the truth-level differential production rate and all identifiable spin coefficients, whereas the usual
flat average over kinematic folds gives significantly biased
reconstructions. 
When a nontrivial null space is present, the reconstructed identifiable subspace together with positivity yields controlled ranges for concurrence and the CHSH parameter.
\end{abstract}

\maketitle
\newpage

\begingroup
\renewcommand{\addcontentsline}[3]{}
\section{Introduction}
\endgroup

Spin information at colliders is inferred from the momentum distributions
of decay products. Particle decay therefore acts as a generalized spin
measurement: the full decay phase space defines a continuous
positive-operator-valued measure (POVM) on the parent spin density
matrix~\cite{Khan:2020seu,Ashby-Pickering:2022umy,Wu:2024mtj,Ai:2025wnt}.
When the decay kinematics can be
fully reconstructed, spin-analyzing angular distributions provide a
direct, data-driven reconstruction of the production spin density matrix,
without assuming a theoretical production template~\cite{Afik:2020onf,Ashby-Pickering:2022umy,Barr:2024djo}.
Related ideas have been applied to top-quark pairs, electroweak dibosons,
tau pairs, and baryon--antibaryon systems at $e^+e^-$ colliders, using
measured decay kinematics together with either tomographic reconstruction
or process-specific spin-density-matrix inputs~\cite{Bernreuther:2015yna,Afik:2020onf,Afik:2022kwm,Aoude:2022imd,Han:2023fci,Ehataht:2023zzt,BESIII:2018cnd,Wu:2024asu,Fabbrichesi:2023cev, BESIII:2025vsr, 
Han:2025ewp}.

Invisible neutrinos can turn this otherwise direct reconstruction into an
ambiguous inverse problem. They arise unavoidably in tau decays and
frequently in clean leptonic or semileptonic channels of heavy particles,
such as top quarks.
Existing analyses therefore adopt practical prescriptions.
Belle flat-averaged the twofold kinematic solutions in optimal-observable
searches for the tau electric dipole moment~\cite{Belle:2002nla,Belle:2021ybo}.
Recent tau-tomography studies have selected or weighted candidate
solutions using auxiliary tracking or vertex information
~\cite{Altakach:2022ywa,Ehataht:2023zzt,Jeans:2026eys}, or retained both
branches in a production-template analysis~\cite{Zhou:2026poo}.
At low-energy STCF, the unresolved twofold ambiguity remains the main
limitation of tau tomography~\cite{Li:2026wxi}.
Dileptonic top reconstructions likewise encounter multiple neutrino solutions that
must be selected or weighted in subsequent analyses~\cite{Sonnenschein:2006pq,Sonnenschein:2006ud,Betchart:2013nba}.
This limitation is reflected in top-pair spin and quantum measurements:
ATLAS and CMS measured the entanglement-sensitive observable $D$ in
dileptonic events rather than reconstructing the full density
matrix~\cite{ATLAS:2023fsd, CMS:2024pts}, while CMS later reconstructed
all polarization and spin-correlation coefficients in semileptonic events
after choosing a kinematic solution~\cite{CMS:2024zkc}.
These strategies are useful, but a systematic, production-model-independent
reconstruction of the density matrix under missing-neutrino ambiguities is
still lacking.

In this work, we formulate collider spin tomography with invisible
neutrinos as a coarse-grained version~\cite{busch1996quantum} 
of the continuous decay POVM.  
Missing neutrinos map the ideal measurement on the full decay kinematics to an
effective measurement on the visible momenta $\Phi$.  
The information content is determined by the kernel of the resulting visible-data map. 
A spin-density-matrix direction is lost only if a nonzero variation along
that direction leaves the observed distribution unchanged for all visible
phase-space points $\Phi$.  
This variational null-space criterion is the continuum analogue of the rank condition for informational completeness in finite-dimensional quantum-state tomography~\cite{2004JOptB...6S.487D, 2013CMaPh.318..355H}, and shows that missing-neutrino ambiguities do not by themselves imply information loss.

We first apply this criterion to
$e^+e^-\to\tau^+\tau^-$ with $\tau^\pm\to\pi^\pm\nu$,
a minimal channel with a twofold kinematic ambiguity and direct relevance
to tau-pair measurements at Belle~II~\cite{Belle-II:2018jsg}, BESIII~\cite{BESIII:2014srs, BESIII:2020nme} and STCF~\cite{Achasov:2023gey}.
In this case, the twofold solutions, the phase-space Jacobian, and the
spin-dependent tau decay distribution can be treated analytically.  
We find a single null direction, $C_{nr}(\hat k)-C_{rn}(\hat k)$.  
The twofold ambiguity therefore does not preclude tomography: the differential production rate and the remaining fourteen spin coefficients are all identifiable from the visible pion distribution.
The lost direction arises from a
non-generic cancellation among the twofold kinematics, the Jacobian
factor, and the decay matrix element.  The information loss in this
channel is therefore specific and controlled, rather than a generic
consequence of neutrinos.

To reconstruct these identifiable quantities, we introduce a self-consistent fixed-point unfolding of the invisible kinematic ambiguity.  The
allowed kinematic folds of each event should not be weighted uniformly: their
weights depend on the phase-space Jacobian and the spin-dependent
differential probability at each fold.  Since this differential
probability itself depends on the unknown production spin density matrix, the
reconstruction is circular.  
We solve this circularity by a fixed-point iteration using
only the observed events and their allowed kinematic solutions, without
assuming any theoretical template for the production density matrix.

We validate the method with simulated tau-pair events.  In the Standard Model (SM) sample, the flat $50/50$ average over the two kinematic folds is visibly biased already in the normalized differential production cross section, as well as in spin observables.  The self-consistent iteration corrects these biases and reproduces the truth values on the identifiable subspace.  
The same closure behavior in anomalous tau-dipole benchmarks confirms that the reconstruction is not an SM template fit.
Although one antisymmetric spin-correlation direction is
not identifiable in the tau-pair case, the reconstructed identifiable
subspace together with density-matrix positivity still yields controlled
ranges for concurrence and CHSH parameters.  
These results separate kinematic ambiguity from information loss and provide a practical, data-driven route to production density matrix reconstruction and
quantum measurements in collider processes with missing energy.

\vspace{20 pt}
\begingroup
\renewcommand{\addcontentsline}[3]{}
\section{Null-space criterion}
\endgroup

We first formulate collider spin tomography as an information-completeness
problem for a coarse-grained continuous POVM.  For a pair of spin-$1/2$
unstable particles, the differential production density matrix at a fixed production
direction $\hat k$ can be written as
\begin{align}
\rho(\hat k)=
\frac{1}{4} \frac{\mathrm d\sigma}{ \sigma \mathrm d \hat{k}  }
&\left[
\mathbb{I}_4
+B_i^+(\hat k)\,\sigma_i\otimes\mathbb{I}_2
+B_i^-(\hat k)\,\mathbb{I}_2\otimes\sigma_i + C_{ij}(\hat k)\,\sigma_i\otimes\sigma_j
\right],
\label{eq:rho-spin}
\end{align}
where ${\mathrm d\sigma}/{(\sigma \, \mathrm d \hat{k} ) }$  is the normalized production distribution,
$B_i^\pm$ are the polarizations, $C_{ij}$ is the spin-correlation
matrix, with $\int {\rm d}\hat{k} \, {\rm Tr}[\rho(\hat{k})]=1$.  We use the standard helicity basis $\{\hat{\mathbf{n}},\hat{\mathbf{r}},\hat{\mathbf{k}}\}$ in the center-of-mass frame. 

If the full decay kinematics $\phi$ are reconstructed, the normalized
distribution is a continuous POVM on the parent spin state~\cite{Ai:2025wnt},
\begin{equation}
f(\phi) \equiv \frac{\rm d\sigma}{\sigma {\rm d\phi}}=
{\rm Tr}\left[\rho(\hat k)E(\phi)\right],
\label{eq:ideal-povm-short}
\end{equation}
where $E(\phi)$ denotes the POVM  associated with the decay configuration.
With invisible particles, experiments observe only the visible variables
$\Phi$.  The full kinematics $\phi$ determine $\Phi$ uniquely, but the inverse
map is generally multi-valued: a given $\Phi$ can correspond to several
allowed configurations $\phi^{t}(\Phi)$.
The visible differential distribution is therefore a linear functional of $\rho$,
\begin{equation}
g(\Phi)\equiv \frac{\mathrm d\sigma}{\sigma \mathrm d\Phi}\equiv \mathcal M[\rho](\Phi).
\label{eq:coarse-map-short}
\end{equation}
The relevant question is not whether neutrinos are missing, nor how many
kinematic solutions exist, but whether this coarse-grained map has a
nontrivial kernel.  An infinitesimal deformation of $\rho$ is invisible if
\begin{equation}
\delta g(\Phi)=\mathcal M[\delta\rho](\Phi)=0,
\quad \text{for all visible  } \Phi .
\label{eq:null-criterion-short}
\end{equation}
Information is lost only for variations $\delta\rho$ that satisfy $\delta g = 0$ for every visible configuration $\Phi$. 
If the only solution in the chosen parameter space
is $\delta\rho=0$, the measurement is informationally complete on that
space.

For finite-fold missing-neutrino ambiguities, the visible distribution is
given by
$g(\Phi)=\sum_t J^{t}(\Phi)f(\phi^{t}(\Phi))$, where
$J^{t}(\Phi)
=\left|\det\frac{\partial \phi^{t}}{\partial \Phi} \right|$
is the Jacobian of the $t$-th fold.  
Thus the coarse-grained POVM takes the concrete form
\begin{align}
g(\Phi)
=\sum_{t} J^{t}\,
{\rm Tr}\left[
\rho(\hat k^{t})E(\phi^{t})
\right] \equiv 
\int {\rm d}\hat{k}\,  {\rm Tr}\left[ \rho(\hat k) \overline{E} \right],
\label{eq:fold-map-short}
\end{align}
where $\overline E(\Phi)=\sum_{t} J^{t}(\Phi)\,E(\phi^{t})\delta(\hat{k}-\hat{k}^{t})$ is the coarse-grained POVM density associated with the map $\mathcal M$. 
Nonzero solutions of Eq.~\eqref{eq:null-criterion-short} are null modes of the
coarse-grained POVM.

We now apply this criterion to
$e^+e^-\to\tau^+\tau^-$ with $\tau^\pm\to\pi^\pm\nu$ in the
$\tau^+\tau^-$ center-of-mass frame.  The full variables are
$\phi=\{\hat k,\hat q_+,\hat q_-\}$, with $\hat q_\pm$ the pion directions
in the corresponding tau rest frames, while the visible variables are the
two pion momenta $\Phi=\{\vec p_+,\vec p_-\}$.  For each visible event
$\Phi$, there are two kinematically allowed tau-direction solutions
$\hat k^a$ and $\hat k^b$, which are mirror images with respect to the
pion plane spanned by $\vec{p}_\pm$.

Substituting this twofold geometry into
Eq.~\eqref{eq:null-criterion-short}, and solving the null-space
condition explicitly, shows that the kernel is one-dimensional.  Every
null deformation is proportional to a single mode,
\begin{equation}
\delta C_{nr}(\hat k)
=-\delta C_{rn}(\hat k)=
d_{\rm null} \times
\left(\frac{\mathrm d\sigma}{\sigma\,\mathrm d\hat{k}}\right)^{-1},
\label{eq:tau-null-short}
\end{equation}
with all other density-matrix components unchanged.  Here $d_{\rm null}$
is an arbitrary constant, independent of $\hat k$; see Appendix.  Thus, the visible pion distribution is blind only to the antisymmetric transverse spin correlation, $C_{nr}^{-}\equiv C_{nr}-C_{rn}$.
The remaining production distribution, polarization and spin-correlation components therefore define the experimentally identifiable subspace.  
This conclusion is fixed by the kinematic coarse graining and the known
decay POVM, and does not use any theoretical information from
$\rho$.

\vspace{20 pt}
\begingroup
\renewcommand{\addcontentsline}[3]{}
\section{Self-consistent unfolding}
\endgroup

We now describe how to reconstruct this identifiable subspace in practice. 
For each visible event, the unresolved fold is a latent variable whose weight depends on the unknown $\rho$.  
Using the fold solutions, we assign the fold weight
\begin{equation}
\omega_t(\Phi;\mathbf x)
=
\frac{
J^t(\Phi)\,
f\left(\phi^{t}(\Phi);\mathbf x\right)
}{
\sum_s
J^s(\Phi)\,
f\left(\phi^{s}(\Phi);\mathbf x\right)
}\,,
\label{eq:fold-weight-1}
\end{equation}
where the vector $\mathbf x =\{\frac{\mathrm d\sigma}{ \sigma \mathrm d \hat{k}  }|_{\alpha},B^+_{i,\alpha},B^-_{i,\alpha},
C_{ij,\alpha}\}$ denotes the 16 functional parameters of $\rho(\hat k)$, with
the production direction $\hat k$ replaced by the bin label $\alpha \in \{1, \ldots,N_{\rm bin}\}$.
Equation~\eqref{eq:fold-weight-1} exposes the central circularity:
the fold weights depend on the unknown density matrix, but the
density-matrix reconstruction itself depends on these fold weights.

Choose full-kinematic observables $O_m(\phi)$ satisfying
$\langle O_m\rangle_\phi=p_m x_m$, where
$m\in{1,\ldots,16N_{\rm bin}}$ labels the components of $\mathbf x$
and the normalization factors $p_m$ are defined in Appendix.
For visible $\Phi$, we define the corresponding fold-weighted observables as
\begin{equation}
\widetilde O_m(\Phi;\mathbf x)
=
\sum_{t}
\omega_t(\Phi;\mathbf x)\,
O_m\left(\phi^{t}(\Phi)\right).
\label{eq:weighted-observable}
\end{equation}
When the input $\mathbf x$ equals the true parameters of $\rho$,
these observables have the same expectation values as the full-kinematic
ones, $\langle \widetilde O_m\rangle_\Phi=p_m x_m$.  
This defines a data-driven update map
$T_m(\mathbf x)\equiv
N_{\rm evt}^{-1}\sum_{\rm events}\widetilde O_m(\Phi_{\rm evt};\mathbf x)$.
Fixed points of the iteration satisfy
\begin{equation}
\mathbf x_*=T(\mathbf x_*).
\label{eq:self-consistency}
\end{equation}

In the infinite-statistics limit, the truth together with arbitrary
displacements along the null direction forms a continuous family of
fixed points:
\begin{equation}
\mathbf x_*(d_{\rm null}) =
\mathbf x_{\rm truth} +
d_{\rm null} \, \mathbf n_{\rm null},
\label{eq:null-fixed-point-family-1}
\end{equation}
where $\mathbf n_{\rm null}$ denotes the direction specified in
Eq.~\eqref{eq:tau-null-short}. Each member of this family produces the
same visible distribution and satisfies Eq.~\eqref{eq:self-consistency},
provided that the corresponding $\rho(\hat{k})$ remains physical.

We solve Eq.~\eqref{eq:self-consistency} by iteration,
$\mathbf x^{(n)}=T(\mathbf x^{(n-1)})$.  
At each step, the trial $\rho$ assigns fold weights to the
kinematic solutions, and the event average of the fold-weighted
observables gives the updated $\rho$ parameters.  
Neither the production Lagrangian nor any theoretical ansatz for $\rho$ is used in this unfolding.
Local convergence toward the fixed-point family is governed by the
Jacobian $(D_T)_{ml}=\partial T_m/\partial x_l$. Numerically, for our
benchmarks, its restriction to the identifiable subspace has spectral
radius below unity, implying local contraction
\cite{ortega1970iterative}, while the exact null direction analytically
has unit eigenvalue; see Appendix. Accordingly, the
identifiable components rapidly approach the truth, 
whereas $C_{nr}^{-}$ approaches an initialization- and
finite-sample-dependent member of the fixed-point family.

This procedure differs from simple prescriptions for the twofold
ambiguity.  A flat average sets $\omega_t=1/N_{\rm fold}$, while a
single-solution prescription sets one branch weight to unity using an
external kinematic rule.  
Both choices impose external assumptions on the unobserved kinematics.  
In contrast, Eq.~\eqref{eq:fold-weight-1} determines the fold
probabilities from the density matrix reconstructed from the data itself.

\begin{figure}[htb]
\centering
\includegraphics[width=\columnwidth]{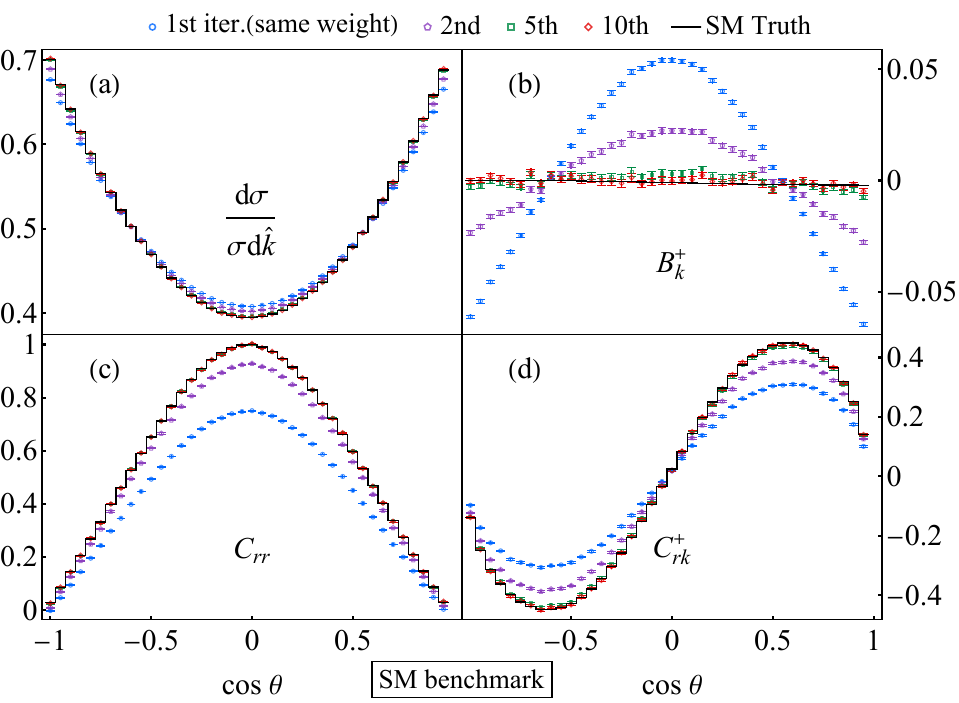}
\caption{
Iterative reconstruction of representative binned density matrix
parameters in the SM
$e^+e^-\to\tau^+\tau^-\to \pi^+\pi^-+\nu\bar\nu$ sample, including the
normalized production-angle distribution
${\rm d}\sigma/(\sigma\,{\rm d}\hat k)$, the polarization component
$B_{k}^{+}$, and the spin-correlation components $C_{rr}$ and
$C_{rk}^+\equiv C_{rk}+C_{kr}$. 
The truth distribution is compared with the 1st, 2nd, 5th,
and 10th iterations, where the error bars show statistical uncertainties.
The 1st iteration corresponds to the flat
$50/50$ average over the two kinematic folds.  The iterations approach
the truth values, correcting the flat-average bias for the 
identifiable components.
}
\label{fig:tau-iteration}
\end{figure}

\vspace{20 pt}

\begingroup
\renewcommand{\addcontentsline}[3]{}
\section{Numerical results}
\endgroup

We now test the self-consistent unfolding with simulated
$e^+e^-\to\tau^+\tau^-\to\pi^+\pi^-+\nu\bar\nu$ events generated at the
Belle~II center-of-mass energy, $\sqrt{s}=10.58~{\rm GeV}$.
The closure tests use $10^8$ generated signal events, corresponding to
about $9.3~{\rm ab}^{-1}$ before detector acceptance and efficiency
effects in this exclusive $\pi^+\pi^-$ channel.
The reconstruction uses only the observed pion momenta and the two allowed
kinematic solutions for each event.  The production matrix element, the
generator-level spin density matrix, and the truth assignment of the
kinematic fold are used only for validation, not as inputs to the
unfolding.

Figure~\ref{fig:tau-iteration} shows the closure test for the tree-level
SM baseline, with no anomalous $\tau$-dipole interaction included.
We initialize the iteration with a spin-independent trial
density matrix.  In the pion channel, the first update is equivalent to a
flat $50/50$ average over the two kinematic folds.  This flat average is
visibly biased not only for spin observables, but also for the normalized
production direction ${\rm d}\sigma/(\sigma\,{\rm d}\hat k)$.
After self-consistent updates, the reconstructed identifiable
components move rapidly toward the generator truth values;
the 5th and 10th iterations are already very close to each other.  
This demonstrates that the data-driven fold weights correct
the bias introduced by a purely flat treatment of the twofold ambiguity.
The nonzero SM truth value of $B_k^+$ reflects the parity-violating
$\gamma$--$Z$ interference contribution.

\begin{figure}[htb]
\centering
\includegraphics[width=\columnwidth]{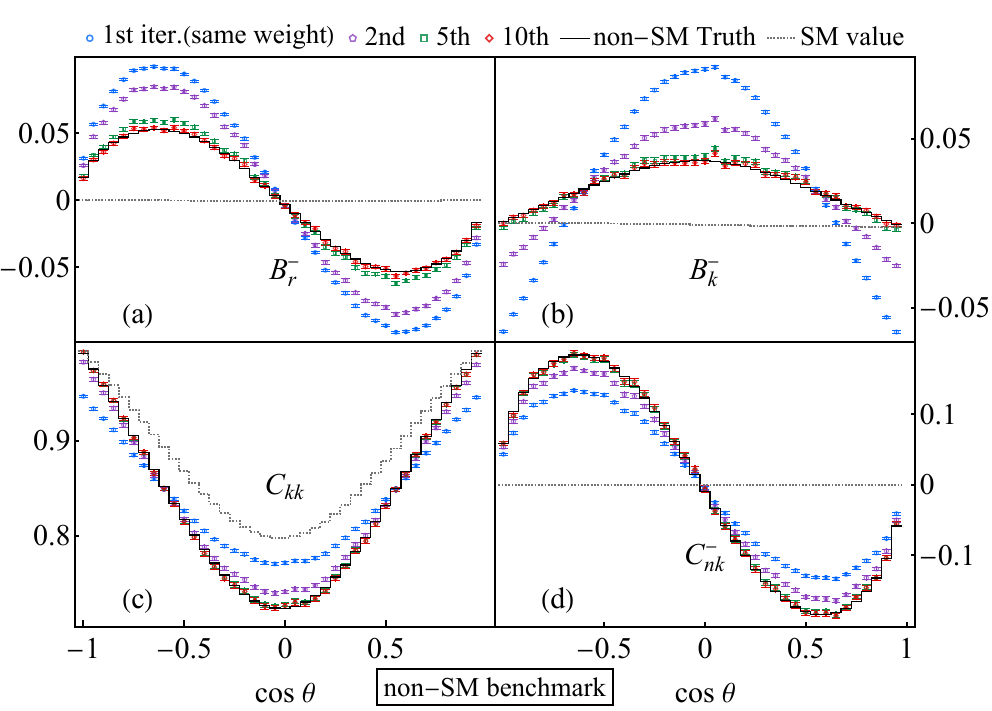}
\caption{
Same as Fig.~\ref{fig:tau-iteration}, but for the non-SM anomalous tau-dipole
benchmark, showing $B_{r}^{-}$, $B_{k}^{-}$,
$C_{kk}$, and $C_{nk}^{-}\equiv C_{nk}-C_{kn} $.}
\label{fig:tau-iteration-nonSM}
\end{figure}

We further test a non-SM spin structure using a deliberately large
anomalous tau-dipole benchmark,
$a_\tau = 0.01 - 0.02i$ and
$d_\tau  (2m_\tau/e) = -0.05 + 0.03i$, chosen as a stress test of the
template independence of the reconstruction. As shown in
Fig.~\ref{fig:tau-iteration-nonSM}, the flat $50/50$ average is again
biased, while the self-consistent iterations converge toward the
non-SM truth values for the displayed identifiable spin observables.
This confirms that the method is not a template fit to the SM
prediction. 
For the null component $C_{nr}^-$, runs with different initializations
approach different limiting values, whose displacements from the truth
agree with Eq.~\eqref{eq:tau-null-short} within finite-sample
fluctuations, providing numerical evidence for the fixed-point family
in Eq.~\eqref{eq:null-fixed-point-family-1}. The remaining SM and non-SM reconstruction results are presented in Appendix.

\begin{figure}[htb]
\centering
\includegraphics[width=0.95\columnwidth]{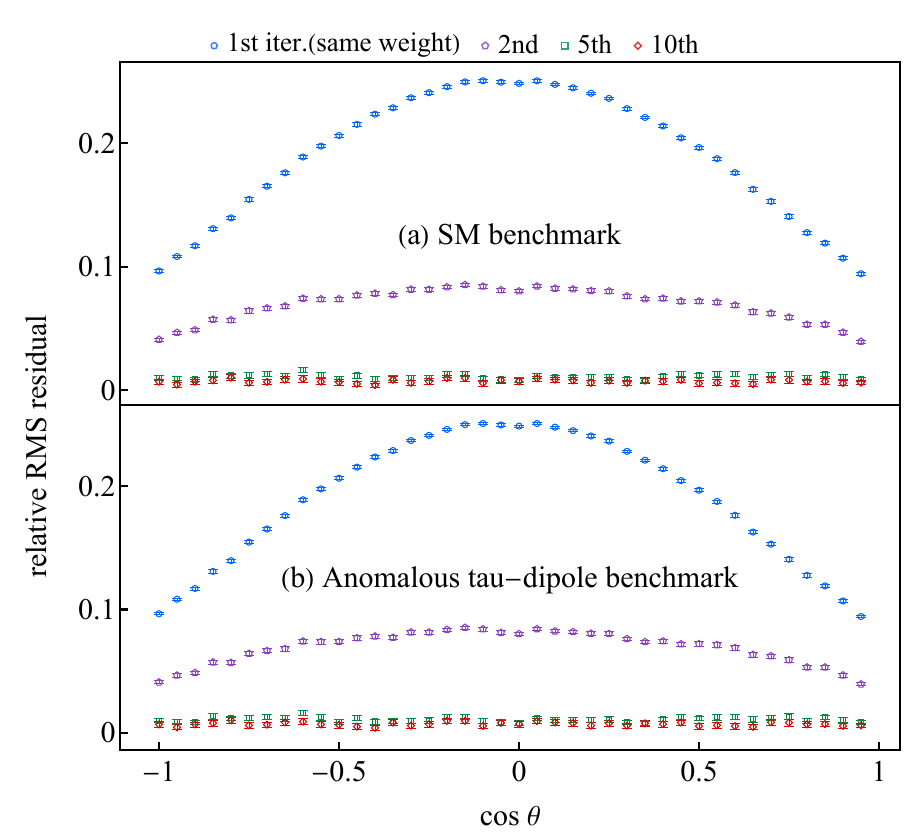}
\caption{
The relative RMS for the identifiable
spin-density-matrix components.  
The residual is computed over all identifiable binned parameters,
including the normalized production-angle distribution, polarizations,
and spin-correlation components, except the null combination $C_{nr}^-$.
(a) SM $\tau\bar{\tau}$ sample.  
(b) Anomalous tau-dipole benchmark.  
Both samples show a clear reduction relative to the flat average.
}
\label{fig:tau-residuals}
\end{figure}

To quantify the reconstruction on the identifiable subspace, we define a relative RMS residual over the binned $\mathbf{x}$ parameters,
excluding the null space,
\begin{equation}
{\rm rRMS}^{(n)}=
\left[
\frac{
\sum_{m \neq{\rm null}}
\left(x^{(n)}_{m}-x^{\rm truth}_{m}\right)^2
}{
\sum_{m \neq{\rm null}}
\left(x^{\rm truth}_{m}\right)^2
}
\right]^{1/2},
\label{eq:global-residual}
\end{equation}
where $n$ denotes the iteration number and $m$ labels the $\mathbf{x}$ components.  Figure~\ref{fig:tau-residuals} shows that the residual
decreases rapidly with iteration in both the SM and anomalous-dipole
samples.  
The flat average gives the largest residual, while 
the self-consistent iterations approach the truth on the identifiable
subspace.

\begin{figure}[htb]
\centering
\includegraphics[width=0.95\columnwidth]{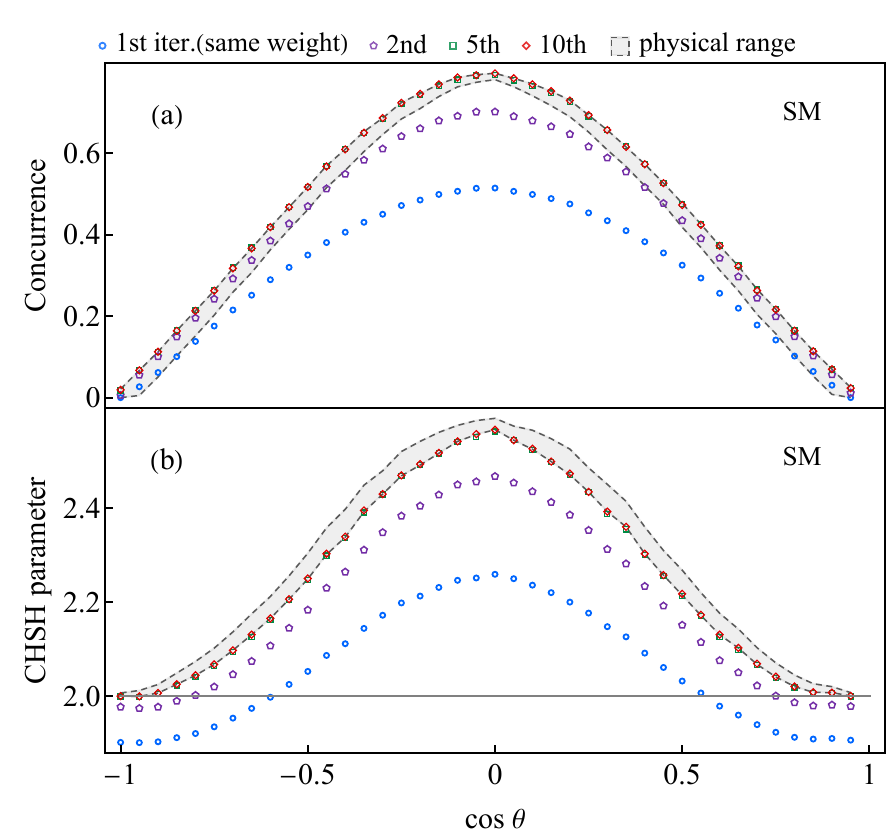}
\caption{
Quantum observables reconstructed from the SM tau-pair
sample.  The unmeasured null component
$C_{nr,\alpha}^-$ is varied over the range allowed by positivity
$\rho_\alpha \succeq 0$, while the identifiable components are fixed to
their reconstructed values.  The resulting intervals are shown for
(a) Concurrence and (b) the CHSH parameter.  
The positivity-constrained intervals are consistent with the truth,
whereas the flat-average method underestimates the displayed
quantum observables.
}
\label{fig:tau-quantum-observables}
\end{figure}

The reconstructed identifiable components can be used to constrain
nonlinear quantum observables, but the null direction must be treated
carefully.
We do not fix the unmeasured component
$C_{nr,\alpha}^-$ using the SM prediction, or any
other theoretical template for $\rho$. 
Instead, for each production-angle bin we keep the reconstructed identifiable
components fixed and vary the null component over all values compatible
with positivity, $\rho_\alpha \succeq 0$ \cite{bengtsson2007geometry}.  This gives allowed intervals
\begin{equation}
\mathcal C_{\min,\alpha}\le \mathcal C_\alpha\le
\mathcal C_{\max,\alpha},
\qquad
\mathcal B_{\min,\alpha}\le \mathcal B_\alpha\le
\mathcal B_{\max,\alpha},
\label{eq:quantum-observable-ranges}
\end{equation}
for the Concurrence $\mathcal C_\alpha$ \cite{Wootters:1997id} and the CHSH parameter
$\mathcal B_\alpha$ \cite{Bell:1964kc, Clauser:1969ny, Horodecki:1995nsk}.  
If $\mathcal C_{\min,\alpha}>0$, the visible distribution supports a
model-independent identification of entanglement.  Similarly, a
positivity-constrained lower bound $\mathcal B_{\min,\alpha}>2$ would
establish a CHSH violation, provided that its excess above 2 is robust
against the numerical positivity tolerance and experimental
uncertainties.
These results should
be interpreted as statistical closure tests of the tomographic method,
rather than as a full experimental sensitivity projection.
Figure~\ref{fig:tau-quantum-observables} shows the reconstructed
concurrence and CHSH intervals in the SM sample.  
The flat average can distort these nonlinear observables because it biases the
underlying spin-density-matrix components.  The self-consistent
reconstruction gives intervals consistent with the truth expectation and
retains strong sensitivity to both entanglement and CHSH violation.  Thus
the single null direction does not prevent collider spin tomography from
accessing quantum correlations; it only requires that observables
depending on the null component be reported as positivity-constrained
ranges rather than as unique numbers.

\vspace{20 pt}
\begingroup
\renewcommand{\addcontentsline}[3]{}
\section{Conclusions}
\endgroup

Collider spin tomography with invisible particles is an information 
completeness problem for a coarse-grained continuous POVM.  Missing
neutrinos are not a fundamental obstruction by themselves: spin
information is lost only along the kernel of the corresponding visible-
data map.  In
$e^+e^-\to\tau^+\tau^-$ with $\tau^\pm \to \pi^\pm+\nu$, this kernel is
one-dimensional and corresponds to the antisymmetric transverse
correlation $C_{nr}^{-}$.  All remaining polarization
and spin-correlation components are identifiable from the visible pion
distribution.

We have developed a self-consistent unfolding method that reconstructs
the identifiable components, including the normalized production-angle
distribution ${\rm d}\sigma/(\sigma\,{\rm d}\hat k)$, the polarizations
$B_i^\pm$, and the identifiable spin-correlation coefficients $C_{ij}$, without using
a theoretical production template.  
The fold weights are determined by the density matrix
reconstructed from the data itself, rather than by a flat average or an
external kinematic choice.  
In the SM tau-pair sample, the flat $50/50$
average is visibly biased even for the normalized production-angle distribution, while the self-consistent iteration corrects this bias and reproduces the
truth-level parameters on the identifiable subspace.  The same
closure behavior in anomalous tau-dipole samples shows that the
reconstruction is not tied to an SM production template.  
The reconstructed components also determine positivity-constrained
intervals for nonlinear quantum observables, such as concurrence and the
CHSH parameter, allowing entanglement and CHSH violation to be tested
without assuming the unmeasured null component.
The same framework also extends to multi-hadron tau decays, such as
$\tau\to\rho\nu\to\pi\pi^0\nu$ and $\tau\to a_1\nu\to 3\pi\nu$, where the
richer decay POVMs and visible response maps are most naturally
constructed numerically. 
In addition, this framework is also applicable to other processes with
multiple invisible particles, including fully leptonic $WW$ and
dileptonic $t\bar{t}$ production.
More broadly, our results turn missing energy from a kinematic obstacle
into a question of POVM identifiability, providing a practical route to
model-independent spin-density-matrix reconstruction and quantum
measurements in collider processes with invisible particles.

\vspace{20 pt}

\begingroup
\renewcommand{\addcontentsline}[3]{}
\section*{Acknowledgements}
\endgroup
The work of J.L. is supported by the National Science Foundation of China under Grant No.~12235001, No.~12475103 and State Key Laboratory of Nuclear Physics and Technology under Grant No. NPT2025ZX11.
The work of X.P.W. is supported by the National Science Foundation of China under Grant No. 12375095, and the Fundamental Research Funds for the Central Universities. 
The authors gratefully acknowledge the valuable discussions and insights provided by the members of the Collaboration of Precision Testing and New Physics.

\clearpage

\onecolumngrid

\appendix

\begin{center}
{\large \bf Appendix \\
Collider Spin Tomography with Missing Neutrinos\\
}

\end{center}

This Appendix provides mathematical proofs, detailed
derivations, and extended numerical studies supporting the conclusions
of the main text.

\tableofcontents

\section{Coarse-grained POVM and Null-space Criterion}
\label{sec:app-POVM-null}

In this section, we present a theoretical analysis of informational completeness in tomography with multifold ambiguities.
We first describe the information-theoretic structure of collider spin tomography.
Consider a pair of unstable spin-$1/2$ particles, such as $\tau^+\tau^-$ or $t\bar t$, whose differential production spin state is described by a density matrix $\rho(\hat k)$, where $\hat k$ denotes the production direction in the center-of-mass frame.
The differential production density matrix can be parameterized as
\begin{equation}
    \rho(\hat k)
    =\mathcal{R}(\hat k)\frac{1}{4}\bigg[ \mathbb{I}_2\otimes\mathbb{I}_2 +\sum_{i}B_i^+(\hat k)(\sigma_i\otimes \mathbb{I}_2)
    +\sum_{j}B_j^-(\hat k)(\mathbb{I}_2\otimes \sigma_j )
    +\sum_{i,j} C_{ij}(\hat k)(\sigma_i\otimes\sigma_j) \bigg],
    \label{eq:app-rho-spin}
\end{equation}
where \(i,j\) represent the three mutually orthogonal directions in three-dimensional space, \(\sigma_i\) denote the Pauli matrices, and \(\mathbb{I}_2\) is the \(2\times2\) identity matrix.
Here
\begin{equation}
\mathcal{R}(\hat k)\equiv
\frac{\mathrm d\sigma}{\sigma\,\mathrm d\hat k}
=\mathrm{Tr}\left[\rho(\hat k)\right]
\end{equation}
is the normalized production-angle distribution, satisfying
$\int\mathrm d\hat k\,\mathcal R(\hat k)=1$.
The polarization vectors and spin-correlation matrix are defined by
\begin{align}
B_i^+(\hat k)
=
\frac{\mathrm{Tr}\left[
\rho(\hat k)(\sigma_i\otimes\mathbb{I}_2)
\right]}{\mathcal R(\hat k)}, \quad
B_j^-(\hat k)
=
\frac{\mathrm{Tr}\left[
\rho(\hat k)(\mathbb{I}_2\otimes\sigma_j)
\right]}{\mathcal R(\hat k)}, \quad
C_{ij}(\hat k)
=
\frac{\mathrm{Tr}\left[
\rho(\hat k)(\sigma_i\otimes\sigma_j)
\right]}{\mathcal R(\hat k)}.
\label{eq:app-Bi-Cij}
\end{align}

All these coefficients are functions of the production kinematics.
It is convenient to choose the helicity basis
$\{\hat{\mathbf n},\hat{\mathbf r},\hat{\mathbf k}\}$ in the center-of-mass frame, defined by
\begin{equation}
    \hat{\mathbf{n}}=\frac{1}{\sin\theta}(\hat{\mathbf{p}}\times\hat{\mathbf{k}}),\quad \hat{\mathbf{r}}=\frac{1}{\sin\theta}(\hat{\mathbf{p}}-\cos\theta\hat{\mathbf{k}}),
\end{equation}
with \(\hat{\mathbf{k}}\) being the direction of the unstable particle, \(\hat{\mathbf{p}}\) being the direction of the electron beam and \(\theta\) the production angle satisfying \(\cos\theta=\hat{\mathbf{p}}\cdot\hat{\mathbf{k}}\).

We divide the full process into production and decay.
The production process prepares the unstable-particle pair in the spin state $\rho(\hat k)$, while the decay process acts as a generalized measurement of that state.
Let $\phi$ denote the full reconstructible phase-space variables, including the production direction $\hat k$ and the measured decay kinematics.
If the full decay kinematics are reconstructed, the normalized differential distribution can be written as
\begin{equation}
f(\phi)\equiv
\frac{\mathrm d\sigma}{\sigma \,\mathrm d\phi} =
\mathrm{Tr}\left[
\rho(\hat k)E(\phi)
\right],
\label{eq:app-ideal-povm}
\end{equation}
where $E(\phi)$ is a positive-operator-valued-measure density labeled by the continuous phase-space variables.
It satisfies positivity and completeness,
\begin{equation}
E(\phi)\succeq0,
\qquad
\int \mathrm d\phi_{\rm dec} \,E(\phi)=\mathbb I_4,
\end{equation}
where $\phi_{\rm dec}$ denotes the decay variables at fixed production kinematics. Together with the completeness condition, this fixes the normalization of the POVM to $\mathrm{Tr}[E]=1/(4\pi^2)$.

For two independent spin-$1/2$ decays, the decay variables are $\phi_{\rm dec}=\{\hat q_+,\hat q_-\}$, where $\hat q_+$ and $\hat q_-$ denote the decay directions in the corresponding parent rest frames. The decay POVM then takes the form
\begin{equation}
\begin{gathered}
E(\hat q_+,\hat q_-)
=
D_+(\hat q_+)\otimes D_-(\hat q_-),
\quad 
D_\pm(\hat q_\pm)
=
\frac{1}{4\pi}
\left(
\mathbb I_2+\alpha_\pm\,
\hat q_\pm\cdot\vec\sigma
\right),
\end{gathered}
\end{equation}
where $\alpha_\pm$ are the corresponding spin-analyzing powers.
For the tau-pair pion decay channel, the spin-analyzing power is maximal. 
With the sign conventions used here,
$\alpha_+=-1$ and $\alpha_-=+1$. 
The full-kinematic distribution is therefore
\begin{equation}
f(\phi)
=
\mathcal R(\hat k)\frac{1}{(4\pi)^2}
\left[
1-\hat q_+^iB_i^+(\hat k)
+\hat q_-^jB_j^-(\hat k)
-\hat q_+^i\hat q_-^jC_{ij}(\hat k)
\right].
\label{eq:app-distri}
\end{equation}
When the full kinematics are available, the angular and energy dependence of $f(\phi)$ generally provides independent spin analyzers from which the spin-density-matrix coefficients can be reconstructed.

In collider processes with invisible neutrinos in the final state, only a set of visible variables $\Phi$ can be measured directly. For the tau-pair with $\tau \to \pi \nu$ decay considered here, these variables are the measured pion momenta, $\Phi=\{\vec p_+,\vec p_-\}$. Consequently, the full kinematic variables $\phi$ are generally not uniquely determined by $\Phi$, and the observed distribution constitutes a coarse-grained measurement of the production density matrix,
\begin{equation}
g(\Phi)
\equiv
\frac{\mathrm d\sigma}{\sigma \,\mathrm d\Phi}
=
\mathcal M[\rho](\Phi),
\label{eq:app-measurement-map}
\end{equation}
where $\mathcal M$ is a linear map from the differential density matrix to the visible distribution.
The relevant question is whether this coarse-grained measurement retains complete information about the density matrix.
The answer is determined by the kernel of $\mathcal M$.
An infinitesimal variation $\delta\rho$ is experimentally invisible if it leaves the observed distribution unchanged,
\begin{equation}
\delta g(\Phi)
= \mathcal M[\delta\rho](\Phi) = 0,
\qquad
\text{for all }\Phi.
\label{eq:app-null-space-condition}
\end{equation}
Nonzero solutions of Eq.~\eqref{eq:app-null-space-condition} are null modes of the coarse-grained POVM and correspond to density-matrix directions that cannot be determined from the visible data alone.
If the only solution in the parameter space of interest is $\delta\rho=0$, the measurement is informationally complete on that space.
The criterion in Eq.~\eqref{eq:app-null-space-condition} is general.

We next consider the case in which missing neutrinos lead to a finite number of kinematically allowed solutions, commonly referred to as a multifold ambiguity. Although the full kinematics $\phi$ determine the visible variables $\Phi$ uniquely, the inverse map from $\Phi$ to $\phi$ is generally not one-to-one. For each observed configuration $\Phi$, we denote the allowed solutions by
\begin{equation}
\phi^t(\Phi),
\qquad
t=1,2,\ldots,N_{\rm fold}.
\end{equation}

More generally, the visible distribution can be written in terms of a response kernel,
\begin{equation}
g(\Phi)
=
\int \mathrm d\phi \,
G(\Phi,\phi)f(\phi).
\label{eq:app-response-kernel}
\end{equation}
For a finite-fold ambiguity, the kernel reduces to
\begin{equation}
G(\Phi,\phi)
=
\delta\left(\Phi-\Phi(\phi)\right) =
\sum_{t=1}^{N_{\rm fold}}
J^t(\Phi)\,
\delta\left(\phi-\phi^t(\Phi)\right),
\end{equation}
where
\begin{equation}
J^t(\Phi)
=
\left|
\det\frac{\partial\phi^t}{\partial\Phi}
\right|
=
\left|
\det\frac{\partial\Phi}{\partial\phi^t}
\right|^{-1}
\end{equation}
is the phase-space Jacobian of the $t$-th inverse-map branch. Equation~\eqref{eq:app-response-kernel} therefore becomes
\begin{align}
g(\Phi)=\sum_{t=1}^{N_{\rm fold}}J^t(\Phi)\,f\!\left(\phi^{t}(\Phi)\right)
=\sum_{t=1}^{N_{\mathrm{fold}}}\mathrm{Tr}\left[\rho(\hat k^t)\, J^{t}(\Phi) E(\phi^t)\right]
\equiv
\int\mathrm d\hat k\,
\mathrm{Tr}\left[
\rho(\hat k)\,
\overline E(\Phi)
\right],
\label{eq:app-multifold-distri}
\end{align}
where
\begin{equation}
\overline E(\Phi)
=
\sum_{t=1}^{N_{\rm fold}}
J^t(\Phi) E\left(\phi^t(\Phi) \right)
\delta\left(\hat k-\hat k^t \right)
\end{equation}
is the coarse-grained POVM associated with the map $\mathcal M$.

Thus, in both finite-fold and continuous-latent-variable cases, invisible particles replace the ideal POVM $E(\phi)$ by an effective coarse-grained POVM defined on the visible variables $\Phi$. Information is lost only for variations $\delta\rho$ satisfying $\delta g(\Phi)=0$ for every visible configuration $\Phi$. In the tau-pair pion channel, the ambiguity is twofold, with
$\phi^{a,b}=\{\hat k^{a,b},\hat q_+^{a,b},\hat q_-^{a,b}\}$, and both the kinematic solutions and the corresponding Jacobians can be obtained analytically, as shown in Sec.~\ref{sec:app-twofold-kin}.

A general variation of the differential production density matrix can be expressed as 
\begin{align}
\delta\rho(\hat k)=
\frac{1}{4}
\bigg[
\delta\mathcal R(\hat k)\, \mathbb I_2\otimes\mathbb I_2 +
\delta \left(\mathcal{R}B_i^+(\hat k)\right)\,\sigma_i\otimes\mathbb{I}_2
+\delta \left(\mathcal{R}B_i^-(\hat k)\right)\,\mathbb{I}_2\otimes\sigma_i+\delta\left(\mathcal{R} C_{ij}(\hat k)\right)\,\sigma_i\otimes\sigma_j
\bigg].
\label{eq:app-delta-rho}
\end{align}
This variation induces corresponding changes $\delta f(\phi)$ and
$\delta g(\Phi)$.
In an actual measurement, the production kinematics are binned, and the continuous functions
$\mathcal R(\hat k)$, $B_i^\pm(\hat k)$, and $C_{ij}(\hat k)$ are replaced by a finite set of bin-averaged parameters.
The map $\mathcal M$ is then projected onto a finite-dimensional parameter space, and Eq.~\eqref{eq:app-null-space-condition} reduces to the standard question of whether the resulting response matrix has nontrivial null directions.

For the decay distribution in Eq.~\eqref{eq:app-distri}, the finite-fold null condition takes the explicit form 
\begin{align}
0=\delta g(\Phi)=
\frac{1}{(4\pi)^2}
\sum_{t=1}^{N_{\rm fold}}
J^t(\Phi)
\left[\delta \mathcal R(\hat k^t)-\hat q_+^{i,t}\,\delta \left(\mathcal{R}B_i^+(\hat k^{t})\right)
+\hat q_-^{j,t}\,\delta\left(\mathcal{R}B_j^-(\hat k^{t})\right)-\hat q_+^{i,t}\hat q_-^{j,t}\,\delta\left(\mathcal{R} C_{ij}(\hat k^{t})\right)
\right],
\label{eq:app-null-multifold}
\end{align}
for all visible $\Phi$.
The continuum formulation therefore makes clear that missing neutrinos do not by themselves imply information loss.
Information is lost only when the specific coarse-grained POVM has a nontrivial kernel, namely, when Eq.~\eqref{eq:app-null-multifold} admits a nonzero variation of the differential production density matrix.

\section{Analytic Null Space in the Tau-pair Pion Channel}
\label{sec:app-twofold}

\subsection{Twofold Kinematics}\label{sec:app-twofold-kin}

We consider tau-pair production followed by the decay of each tau into a pion and a neutrino,
\begin{equation}
e^+(p_e)+e^-(-p_e) \rightarrow \tau^+(k)+\tau^-(-k) \rightarrow \pi^+(p_+)+\pi^-(p_-)+ \nu \bar{\nu},
\end{equation}
where the momenta in parentheses are defined in the tau-pair center-of-mass frame.
The six-dimensional visible phase space of the pion momenta is denoted by
$\Phi=\{\vec p_+,\vec p_-\}$, with
$\mathrm d\Phi\equiv \mathrm d^3\vec p_+ \,\mathrm d^3\vec p_-$.
Boosting the pion momenta into the rest frames of their corresponding parent taus gives the decay directions $\hat q_\pm$.
The full kinematic variables can therefore be written as
$\phi=\{\hat k,\hat q_+,\hat q_-\}$.
We use the helicity basis
$\{\hat{\mathbf n},\hat{\mathbf r},\hat{\mathbf k}\}$
in the tau-pair center-of-mass frame.

The invisible neutrinos lead to a twofold ambiguity in the event reconstruction.
Using only the measured pion three-momenta ${\vec p_\pm}$, the parent-tau momentum direction and the pion decay directions in the corresponding tau rest frames,
$\{\hat k,\hat q_+,\hat q_-\}$, cannot be determined uniquely.
Instead, one obtains two kinematically allowed solutions, denoted by $\phi^{a/b}\equiv\{\hat{k}^{a/b}, \hat{q}_{+}^{a/b}, \hat{q}_{-}^{a/b}\}$.
The two solutions for the tau direction are
\begin{align}
    \hat{k}^{(a)}&=u\hat{p}_++v\hat{p}_-+w\frac{\vec{p}_+\times\vec{p}_-}{|\vec{p}_+\times\vec{p}_-|} \label{eq:app-twofold-k1}, \\
    \hat{k}^{(b)}&=u\hat{p}_++v\hat{p}_--w\frac{\vec{p}_+\times\vec{p}_-}{|\vec{p}_+\times\vec{p}_-|}
    \label{eq:app-twofold-k2},
\end{align}
where $\hat p_\pm$ are unit vectors along the pion momenta, and
\begin{align}
    u=& \frac{\cos\theta_++\hat{p}_+\cdot\hat{p}_-\cos\theta_-}{1-(\hat{p}_+\cdot\hat{p}_-)^2},\\
    v=& -\frac{\cos\theta_-+\hat{p}_+\cdot\hat{p}_-\cos\theta_+}{1-(\hat{p}_+\cdot\hat{p}_-)^2},\\
    w=&\sqrt{1-u^2-v^2-2uv(\hat{p}_+\cdot\hat{p}_-)}.
\end{align}

\begin{figure}
    \centering
    \includegraphics[width=0.6\linewidth]{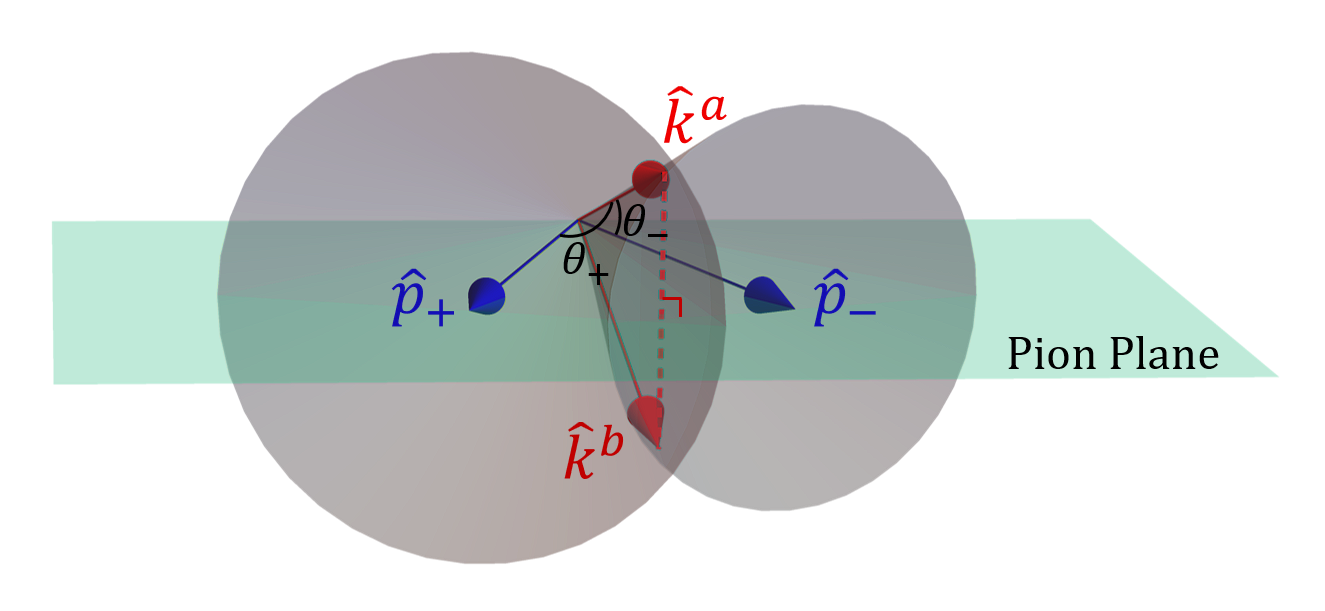}
    \caption{Schematic illustration of the geometric configuration for the twofold kinematics.
    The blue arrows denote the momenta of two measured pions in the tau-pair center-of-mass frame.
    The two allowed solutions to the parent-tau momentum directions in the COM frame, $\hat k_A$ and $\hat k_B$, align with the intersections of two cones centered around $\hat p_\pm$.}
    \label{fig:twofold-schematic}
\end{figure}

Here, the angles $\theta_\pm$ are defined as the angles between the pion momenta $\vec{p}_\pm$ and the corresponding tau momenta. They depend only on the magnitudes of the pion momenta $|\vec p_\pm|$:
\begin{equation}
    \cos\theta_\pm=\frac{2E_\tau E_\pm-m_\pi^2-m_\tau^2}{2|\vec{k}||\vec{p}_\pm|},
\end{equation}
where $E_\pm$ are the energies of $\pi^\pm$, namely,
$E_\pm=\sqrt{m_\pi^2+|\vec p_\pm|^2}$.
For given pion three-momenta in the tau-pair center-of-mass frame, the angles $\theta_\pm$ are fixed. The direction $\hat{k}$ must therefore lie simultaneously on two cones centered around $\vec{p}_\pm$, with half-opening angles $\theta_\pm$.
Consequently, the two allowed solutions for $\hat{k}$ are given by the intersections of these two cones, as shown in Fig.~\ref{fig:twofold-schematic}.
Once a solution for $\hat{k}$ is chosen, the corresponding decay directions $\hat{q}_\pm$ can be computed as
\begin{equation}
    \hat{q}_\pm= \frac{2}{m_\tau^2-m_\pi^2}\bigg[m_\tau\vec{p}_\pm \mp \frac{m_\tau^2+m_\pi^2+2m_\tau E_\pm}{2(E_\tau+m_\tau)}\vec{k}\bigg].
    \label{eq:app-twofold-q}
\end{equation}
Conversely, given the set $\{\hat{k}, \hat{q}_+, \hat{q}_-\}$, the pion momenta $\{\vec{p}_\pm\}$ can be uniquely determined by Lorentz boosting from the corresponding tau rest frames:
\begin{equation}
    \vec{p}_\pm=\frac{m_\tau^2-m_\pi^2}{2m_\tau}\bigg[\hat{q}_\pm+\frac{E_\tau-m_\tau}{m_\tau}(\hat{k}\cdot \hat{q}_\pm)\hat{k}\pm\frac{m_\tau^2+m_\pi^2}{m_\tau^2-m_\pi^2}\frac{\vec{k}}{m_\tau}\bigg].
    \label{eq:app-twofold-p}
\end{equation}

From the geometric structure described above, the exchange of the two branches can be understood as a mirror reflection in the three-dimensional space containing the measured pion momenta.
The reflection operator relating the two branches can therefore be defined as
\begin{align}
\mathsf M_\pi=\mathbb I_3-2\hat n_\pi\hat n_\pi^T,
\end{align}
where $\hat n_\pi=\frac{\vec p_+\times\vec p_-}{|\vec p_+\times\vec p_-|}$ is the unit vector normal to the pion plane.
Applying this reflection operator to one solution gives the other:
\begin{equation}
\hat k^b=\mathsf M_\pi\hat k^a,\qquad
\hat q_\pm^b=\mathsf M_\pi\hat q_\pm^a .
\label{eq:app-mirror}
\end{equation}
The second relation follows from Eq.~\eqref{eq:app-twofold-q}, because
$\mathsf M_\pi\vec p_\pm=\vec p_\pm$ and $E_\pm$ are fixed by the measured pion energies.

The observed probability distribution is
\begin{equation}
    g(\Phi)=\sum_{t=1}^2 J^{t}(\Phi) f(\phi^{t}),
    \label{eq:app-twofold-distri}
\end{equation}
where the $J(\Phi) \equiv \left|\det\!\left(\frac{\partial \phi }{\partial \Phi}\right)\right|= \left|\det\!\left(\frac{\partial \{\hat{k},\hat{q}_{+},\hat{q}_{-} \} }{\partial \{\vec p_{+},\vec p_{-} \} }\right)\right|$ is the Jacobian between \(\phi\) and \(\Phi\).
For the process considered here, the inverse Jacobian $J^{-1}(\Phi)$ can be calculated analytically using Eq.~\eqref{eq:app-twofold-p}:
\begin{align}
    J^{-1}(\Phi)  \equiv \left|\det\!\left(\frac{\partial \{ \vec p_{+},\vec p_{-}\} }{\partial \{ \hat{k},\hat{q}_{+},\hat{q}_{-} \}}\right)\right| =
     \frac{(m_\tau^2-m_\pi^2)^6k^4}{64m_\tau^{10}}\left|\sin\theta_+\sin\theta_-\sin(\phi_--\phi_+)h(\cos\theta_+,\cos\theta_-)\right|,
     \label{eq:app-twofold-jacobi}
\end{align}
where $\theta_\pm$ and $\phi_\pm$ are the polar and azimuthal angles of $\hat{q}_\pm$ in the helicity basis, $E_\tau=\sqrt{s}/2$ is the tau energy,
$k\equiv|\vec{k}|=\sqrt{E_\tau^2-m_\tau^2}$ is the magnitude of the tau momentum, and
$h(\cos\theta_+,\cos\theta_-)$ is a function of $\{\cos\theta_+,\cos\theta_-\}$:
\begin{equation}
    h(\cos\theta_+,\cos\theta_-)\equiv\cos\theta_+\cos\theta_-+\frac{E_\tau(m_\tau^2+m_\pi^2)}{k(m_\tau^2-m_\pi^2)}(\cos\theta_--\cos\theta_+)-\frac{E_\tau^2}{k^2}\bigg(\frac{m_\tau^2+m_\pi^2}{m_\tau^2-m_\pi^2}\bigg)^2.
\end{equation}

Since the mirror reflection is an angular isometry, the branch-exchange Jacobian has unit magnitude.
Equivalently, for this twofold inverse map, the signed visible-map Jacobian changes sign under the mirror reflection, while its absolute value remains unchanged.
Hence,
\begin{equation}
J^a(\Phi)=J^b(\Phi).
\end{equation}
This result can also be checked explicitly using Eq.~\eqref{eq:app-twofold-jacobi}.
Owing to the specific geometric structure of the tau-pair decay, the Jacobian can be written as
\begin{align}\label{eq:app-twofold-jacobi-2}
\left|J^t(\Phi)\right|^{-1}&\propto |\sin\theta_k^{+,t}\sin\theta_k^{-,t}\sin(\phi_k^{+,t}- \phi_k^{-,t})|\cdot h(\hat q_+^t,\hat q_-^t),\nonumber\\
&= |\cos\theta_n^{+,t}\cos\theta_r^{-,t}
-\cos\theta_r^{+,t}\cos\theta_n^{-,t}|\cdot h(\hat q_+^t,\hat q_-^t),
\end{align}
where the function $h(\hat q_+^t,\hat q_-^t)$ takes the same value on the two branches.
Moreover,
\begin{align}
\cos\theta_n^{+,t}\cos\theta_r^{-,t}-\cos\theta_r^{+,t}\cos\theta_n^{-,t}
=(\hat{q}_+^t\times \hat{q}_-^t)\cdot \hat{k}^t\propto
(\vec{p}_+\times \vec{p}_-)\cdot \hat{k}^t,
\end{align}
which immediately verifies the result.

\subsection{Analytic Determination of the Null Space}
\label{sec:app-twofold-null}

The criterion in Eq.~\eqref{eq:app-null-multifold} is general but is usually difficult to solve analytically, as it involves the full kinematic map, the phase-space Jacobian, and the decay matrix elements. The tau-pair pion channel provides a rare analytically tractable example.
We now apply the null-space criterion to
$e^+e^-\to\tau^+\tau^-\to\pi^+\pi^-+\nu\bar\nu$
and show that its twofold ambiguity gives rise to a special antisymmetric null sector.

Heuristically, using Eq.~\eqref{eq:app-twofold-jacobi-2}, one can directly identify a solution by choosing the variations as
\begin{equation}
    \delta \left( \mathcal{R}C_{ij}(\hat{k}) \right)=
    \begin{pmatrix}
        0 & \delta \left( \mathcal{R}C_{nr}^-(\hat{k}) \right) & 0\\
        -\delta \left( \mathcal{R}C_{nr}^-(\hat{k})\right) & 0 & 0\\
        0 & 0 & 0
    \end{pmatrix},
\end{equation}
with all other components set to zero. Equation~\eqref{eq:app-null-multifold} then reduces to
\begin{align}
|J^a|\left(\hat q_+^{n,a}\hat q_-^{r,a}-\hat q_+^{r,a}\hat q_-^{n,a}\right)\,
\delta\left(\mathcal{R}C_{nr}^-(\hat{k}^a)\right)
=-|J^b|\,\left(\hat q_+^{n,b}\hat q_-^{r,b}-\hat q_+^{r,b}\hat q_-^{n,b}\right)
\delta\left(\mathcal{R}C_{nr}^-(\hat{k}^b)\right).
\label{eq:app-null-twofold}
\end{align}

Substituting Eq.~\eqref{eq:app-twofold-jacobi-2}, we find a nonzero solution to Eq.~\eqref{eq:app-null-multifold}, corresponding to the null direction
$\delta\left(\mathcal{R}C_{nr}^-(\hat{k})\right)=\mathrm{const}$.
This suggests that information along this direction is lost because of the missing neutrinos, while the remaining directions may still be identifiable.
We next prove this statement.

For convenience, we define
\begin{align}
\mathcal R(\hat k)=\frac{\mathrm d\sigma}{\sigma\,\mathrm d\hat k},\;
\Delta_0(\hat k)=\delta\mathcal R(\hat k),\;
(\Delta_\pm)_i(\hat k)=\mp\delta\left(\mathcal RB_i^\pm(\hat k)\right),\;
(\Delta_C)_{ij}(\hat k)=-\delta\left(\mathcal RC_{ij}(\hat k)\right),
\end{align}
where $\Delta_\pm$ are three-dimensional vectors and $\Delta_C$ is a $3\times3$ matrix.
All components are expressed in the helicity basis
$\{\hat{\mathbf n},\hat{\mathbf r},\hat{\mathbf k}\}$
associated with the corresponding tau direction.
Using $J^a(\Phi)=J^b(\Phi)$ and dividing out the common nonzero
Jacobian away from the fold-coincidence boundary, the null condition
in Eq.~\eqref{eq:app-null-multifold} becomes 
\begin{equation}
\sum_{t=a,b}
\left[
\Delta_0^t(\hat{k})
+\hat q_+^t\cdot\Delta_+^t(\hat{k})
+\hat q_-^t\cdot\Delta_-^t(\hat{k})
+(\hat q_+^t)_i(\Delta_C^t)_{ij}(\hat{k})(\hat q_-^t)_j
\right]
=0,
\qquad \text{for all visible }\Phi .
\label{eq:app-null-multifold-2}
\end{equation}
The relation at the fold-coincidence boundary is understood by
continuity.

We first continuously approach the configurations where the two folds
coincide. This happens exactly when $\hat k$ lies within the pion plane,
i.e. $\hat n_\pi\cdot\hat k=0$. In this limit,
$\hat q_+$, $\hat q_-$, and $\hat k$ become coplanar, implying
$(\hat q_+\times\hat q_-)\cdot\hat k=0$. Therefore, taking the
coincidence limit of Eq.~\eqref{eq:app-null-multifold-2} leads to the
following necessary condition:
\begin{equation}
\Delta_0(\hat{k})+\hat q_+\cdot\Delta_+(\hat{k})
+\hat q_-\cdot\Delta_-(\hat{k})
+(\hat q_+)_i(\Delta_C)_{ij}(\hat{k})(\hat q_-)_j=0,
\quad\text{for}\quad
(\hat q_+\times\hat q_-)\cdot\hat k=0 .
\label{eq:null-coincidence}
\end{equation}

For a fixed $\hat k$, let $\mathcal{A}$ denote an arbitrary plane
containing $\hat k$. Since $\hat q_+$ and $\hat q_-$ can be independently varied within
$\mathcal A$, the different structures in
Eq.~\eqref{eq:null-coincidence} cannot cancel each other for all
configurations. Therefore, each independent contribution must vanish
separately, leading to
\begin{equation}
\Delta_0=0,\qquad
P_{\mathcal{A}}\Delta_\pm=0,\qquad
P_{\mathcal{A}}\Delta_C P_{\mathcal{A}}=0,
\label{eq:local-null}
\end{equation}
where
\[
P_{\mathcal{A}}=1-\hat n_{\mathcal{A}}\hat n_{\mathcal{A}}^T
\]
denotes the orthogonal projector onto the plane $\mathcal{A}$.

The second constraint implies $\Delta_\pm$ should be perpendicular to $\mathcal{A}$. By allowing $\mathcal{A}$ to vary over all planes containing $\hat k$,
we obtain
\begin{equation}
\Delta_+(\hat k)=\Delta_-(\hat k)=0 .
\end{equation}

The third constraint can be further analyzed by choosing successively
the planes
\[
\operatorname{span}(\hat e_n,\hat e_k),\qquad
\operatorname{span}(\hat e_r,\hat e_k),\qquad
\operatorname{span}((\hat e_n+\hat e_r)/\sqrt2,\hat e_k).
\]
These choices eliminate all components of $\Delta_C$ except for an
antisymmetric pair in the transverse plane. Hence, the only possible
local structure is
\begin{equation}
\Delta_C(\hat k)=d(\hat k)
\begin{pmatrix}
0&1&0\\
-1&0&0\\
0&0&0
\end{pmatrix}_{(n,r,k)} .
\label{eq:local-antisymmetric}
\end{equation}

Substituting this form back into the exact null condition
Eq.~\eqref{eq:app-null-multifold-2} gives
\begin{equation}
d(\hat k^a)=d(\hat k^b).
\label{eq:b-branch}
\end{equation}

For fixed $\hat k^a$, varying the pion plane continuously allows
$\hat k^b$ to cover a connected neighborhood of $\hat k^a$.
Equation~\eqref{eq:b-branch} therefore implies that $d(\hat k)$ is
constant within each such patch. Since the accessible $\hat k$ space
can be covered by overlapping connected patches, $d(\hat k)$ must be
constant over the entire accessible domain.
Denoting this constant by $d_{\rm null}$ according to the sign
convention adopted in the folded null condition, the complete null space
in the helicity basis is one-dimensional:
\begin{equation}
\mathcal R(\hat k)\,\delta C_{nr}(\hat k)
=-\mathcal R(\hat k)\,\delta C_{rn}(\hat k)
=d_{\rm null}=\mathrm{const.},
\label{eq:final-null}
\end{equation}
while all other variations,
including $\delta\mathcal R(\hat k)$,
$\delta B_i^\pm(\hat k)$, and the remaining
$\delta C_{ij}(\hat k)$, vanish. The factor
$\mathcal R(\hat k)$ can be taken outside the variation because
$\delta\mathcal R(\hat k)=0$.

\section{Self-consistent Unfolding in Binned Parameter Space}
\label{sec:app-unfold}

\subsection{Reweighted Observables and Tomographic Estimators}
\label{sec:app-unfold-tomography}

If the ambiguity can be resolved, one can define an observable \(O(\phi)\) as a function of \(\phi\) and calculate its average value over the data set: 
\begin{equation}
    \langle O\rangle=\frac{1}{N_{\rm event}}\sum_{\rm event}O(\phi).
\end{equation}
For a sufficiently large number of events $N_{\rm event}$, this sample average approaches the expectation value with respect to the distribution $f(\phi)$:
\begin{equation}
    \langle O \rangle=\mathbb{E}_\phi[O]\equiv\int \mathrm d\phi\, f(\phi)O(\phi).
    \label{eq:app-exp-value-phi}
\end{equation}
Thus, in the large-sample limit, calculating the average over the data is equivalent to evaluating the integral in Eq.~\eqref{eq:app-exp-value-phi}.

When the ambiguity cannot be resolved, however, an experimentally accessible observable can depend only on the visible variables $\Phi$. Its sample average approaches the expectation value with respect to the visible distribution $g(\Phi)$: 
\begin{equation}
     \mathbb{E}_\Phi[O]\equiv
     \int d\Phi \sum_{t=1}^{N_{\rm fold}} J^{t}(\Phi)f\left(\phi^{t}(\Phi)\right) \times O(\Phi),
\end{equation}
where we have used Eq.~\eqref{eq:app-multifold-distri}.
For an observable $O(\phi)$ defined on the full phase space, we can construct a corresponding observable $\widetilde{O}(\Phi)$ on the visible phase space as 
\begin{equation}
    \widetilde{O}(\Phi) \equiv \frac{\sum_{t}J^{t}(\Phi)f(\phi^{t}(\Phi)) \times O[\phi^{t}(\Phi)]}{\sum_{t}J^{t}(\Phi)f(\phi^{t}(\Phi))}.
    \label{eq:app-obs-Phi}
\end{equation}
Consequently, the expectation value of \(\widetilde{O}(\Phi)\) is equal to the expectation value of \(O(\phi)\):
\begin{align}
\mathbb{E}_\Phi[\widetilde{O}] & =
\int \mathrm d\Phi ~\widetilde{O}(\Phi)  \sum_{t}J^{t}(\Phi)f(\phi^{t}(\Phi))
=\int \mathrm d\Phi \sum_{t}J^{t}(\Phi)f(\phi^{t}(\Phi))O[\phi^{t}(\Phi)] \\
& =\sum_{t}\int\mathrm d\phi^{t} f(\phi^{t})O(\phi^{t}) 
= \int\mathrm d\phi f(\phi)O(\phi)
=\mathbb{E}_\phi[O].
\label{eq:app-exp-value-Phi}
\end{align}
In the following, we focus on spin tomography in the twofold tau-pair process and construct observables whose expectation values determine the parameters $B_i^\pm$ and $C_{ij}$.

As discussed above, \(B_i^\pm\), \(C_{ij}\), and \(\mathrm d\sigma/(\sigma \,\mathrm d\hat k)\) are functions of \(\hat k\).
In collider experiments, these continuous functions cannot be determined point by point, because an infinitesimal phase-space region contains vanishingly few events. 
In practice, we can discretize these continuous functions by dividing the range of \(\hat{k}\) into several bins.  
Since the system possesses rotational symmetry about the incident beam axis, it is sufficient to bin the polar angle \(\theta\) of \(\hat{k}\). After integrating over the azimuthal angle, we use
$\mathrm d\hat k\equiv \mathrm d\cos\theta$ and write
\(
\mathcal R(\hat k)
\equiv
\frac{\mathrm d\sigma}
{\sigma\,\mathrm d\cos\theta}.
\) 
In practice, we divide \(\cos\theta\) evenly into \(N_\text{bin}=40\) bins from $-1$ to $1$, resulting in a bin width of \(w_\text{bin}=\frac{2}{N_\text{bin}}=0.05\) .

Within a specific bin, the functions \(B^\pm_i\) and \(C_{ij}\) take their average values weighted by the differential cross section in that bin, while \(\frac{\mathrm d\sigma}{\sigma \mathrm d\hat{k}} \) is replaced in each bin by its bin-averaged density $\mathcal{R}_\alpha$. The corresponding probability for an event to lie in the $\alpha$th bin is
\begin{equation}
    p_\alpha \equiv \mathcal{R}_\alpha w_{\rm bin}=\int _{\hat{k}\text{ in } \alpha \text{th bin}}  \mathrm d\hat{k} \, \mathcal{R}(\hat k).
\end{equation}
In the \(\alpha\)th bin (\(\alpha=1,2,\ldots,N_\text{bin}\)), the average values of these functions can be calculated as 
\begin{equation}
\begin{aligned}
\mathcal{R}_\alpha &= \frac{1}{w_\text{bin}} \int _{\hat{k}\text{ in } \alpha \text{th bin}}  \mathrm d\hat{k} \, \mathcal{R}(\hat k) ,\\
 B_{i;\alpha}^\pm &=\frac{1}{p_\alpha} \int _{\hat{k}\text{ in } \alpha \text{th bin}} \mathrm d\hat{k} \, \mathcal{R}(\hat k)\, B_{i}^\pm(\hat{k}),\\
C_{ij;\alpha} &= \frac{1}{p_\alpha} \int _{\hat{k}\text{ in } \alpha \text{th bin}} \mathrm d\hat{k} \, \mathcal{R}(\hat k)\, C_{ij}(\hat{k}).
\end{aligned}
\label{eq:app-binned-parameters}
\end{equation}
Using these \(16N_\text{bin}\) parameters, the distribution function \(f(\phi)\) in Eq.~\eqref{eq:app-distri} is now discretized as 
\begin{equation}
    f(\phi_\alpha)=\mathcal{R}_\alpha\frac{1}{(4\pi)^2}\left(1 -\hat{q}_+^iB_{i;\alpha}^++\hat{q}_-^jB_{j;\alpha}^- -\hat{q}_+^i\hat{q}_-^jC_{ij;\alpha}\right).
    \label{eq:app-binned-distri}
\end{equation}
The goal of tomography is to determine these parameters from the observed distribution in experiments. 

When the twofold ambiguity can be resolved, one can access the full-kinematic distribution \(f(\phi)\) in Eq.~\eqref{eq:app-distri}.
To obtain the \(16N_{\mathrm{bin}}\) parameters  defined in Eq.~\eqref{eq:app-binned-parameters}, we can define observables as
\begin{equation}
O_{\mathcal{R}_\alpha}=\frac{\eta(\alpha)}{w_\text{bin}},\qquad
O_{B_{i;\alpha}^\pm}=\mp3\hat{q}_\pm^i\eta(\alpha),\qquad
O_{C_{ij;\alpha}}=-9\hat{q}^i_+\hat{q}^j_-\eta(\alpha),
\label{eq:app-obs-tomo-without}
\end{equation}
where \(\eta(\alpha)\) is the bin-selection function defined by
\begin{equation}
    \eta(\alpha)=\begin{cases}
        1,&\text{if } \hat{k} \text{ in }\alpha\text{th bin},\\
        0,& \text{if } \hat{k} \text{ not in }\alpha\text{th bin}.
    \end{cases}
\end{equation}
The expectation values of observables in Eq.~\eqref{eq:app-obs-tomo-without} are
\begin{equation}
\mathbb{E}_\phi[O_{\mathcal{R}_\alpha}]=\mathcal{R}_\alpha,\qquad
\mathbb{E}_\phi[O_{B_{i;\alpha}^\pm}]=p_\alpha B_{i;\alpha}^\pm,\qquad
\mathbb{E}_\phi[O_{C_{ij;\alpha}}]=p_\alpha C_{ij;\alpha}.
\end{equation}
Therefore, $\mathcal{R}_\alpha$ is obtained directly from the expectation value of $O_{\mathcal{R}_\alpha}$, whereas the normalized polarization and
spin-correlation parameters are obtained from the conditional
expectation values within the $\alpha$th bin:
\begin{equation}
B^\pm_{i;\alpha}= \mathbb{E}_{\phi|\alpha}[O_{B^\pm_{i;\alpha}}]\equiv\frac{\mathbb{E}_\phi[O_{B^\pm_{i;\alpha}}]}{p_\alpha}=\frac{\mathbb{E}_\phi[O_{B^\pm_{i;\alpha}}]}{\mathcal{R}_\alpha w_{\rm bin}}, \qquad
C_{ij;\alpha}=\mathbb{E}_{\phi|\alpha}[O_{C_{ij;\alpha}}] \equiv\frac{\mathbb{E}_\phi[O_{C_{ij;\alpha}}]}
{p_\alpha}=\frac{\mathbb{E}_\phi[O_{C_{ij;\alpha}}]}
{\mathcal{R}_\alpha w_{\rm bin}}.
\end{equation}

When twofold ambiguity cannot be resolved, one can only access the distribution \(g(\Phi) \). 
Using Eqs.~\eqref{eq:app-obs-Phi} and \eqref{eq:app-exp-value-Phi}, we define the corresponding visible observables as
\begin{equation}
\widetilde{O}_{\mathcal{R}_\alpha}(\Phi)
=
\frac{
\sum_{t=1}^2
J^tf(\phi^t)
O^t_{\mathcal{R}_\alpha}
}{
\sum_{t=1}^2
J^tf(\phi^t)
},
\;
\widetilde{O}_{B_{i;\alpha}^\pm}(\Phi)
=
\frac{
\sum_{t=1}^2
J^tf(\phi^t)
O^t_{B_{i;\alpha}^\pm}
}{
\sum_{t=1}^2
J^tf(\phi^t)
},\;
\widetilde{O}_{C_{ij;\alpha}}(\Phi)=\frac{
\sum_{t=1}^2
J^tf(\phi^t)
O^t_{C_{ij;\alpha}}
}{
\sum_{t=1}^2
J^tf(\phi^t)
} \,,
\label{eq:app-obs-tomo-with}
\end{equation}
which satisfies
\begin{equation}
\mathbb{E}_\Phi[\widetilde{O}_{\mathcal{R}_\alpha}]=\mathcal{R}_\alpha,\qquad
\mathbb{E}_\Phi[\widetilde{O}_{B_{i;\alpha}^\pm}]=p_\alpha B_{i;\alpha}^\pm,\qquad
\mathbb{E}_\Phi[\widetilde{O}_{C_{ij;\alpha}}]=p_\alpha C_{ij;\alpha}.
\end{equation}
Here the dependence of $J^t$, $\phi^t$, and $O^t$ on the
observed configuration $\Phi$ is left implicit.
Consequently, the binned parameters are reconstructed as
\begin{equation}
    \mathcal{R}_\alpha=\mathbb{E}_\Phi[\widetilde{O}_{\mathcal{R}_\alpha}],\quad B^\pm_{i;\alpha}=\frac{\mathbb{E}_\Phi[\widetilde{O}_{B^\pm_{i;\alpha}}]}{p_\alpha}=\frac{\mathbb{E}_\Phi[\widetilde{O}_{B^\pm_{i;\alpha}}]}{\mathcal{R}_\alpha w_{\rm bin}}, \quad
C_{ij;\alpha}=\frac{\mathbb{E}_\Phi[\widetilde{O}_{C_{ij;\alpha}}]}
{p_\alpha}=\frac{\mathbb{E}_\Phi[\widetilde{O}_{C_{ij;\alpha}}]}
{\mathcal{R}_\alpha w_{\rm bin}}.
\label{eq:app-self-consistent-equation}
\end{equation}

However, evaluating these observables requires the values of the full-phase-space probability density $f(\phi^t)$ at the allowed folds for each event, and these values themselves depend on \(\mathcal R_\alpha\), \(B_{i;\alpha}^\pm\), and \(C_{ij;\alpha}\).
Thus, Eq.~\eqref{eq:app-self-consistent-equation} is essentially a self-consistent equation \(\mathbf{x}=T(\mathbf{x})\), where \(\mathbf{x}\equiv\{\mathcal{R}_\alpha,B_{i;\alpha}^\pm,C_{ij;\alpha} \}\) is a \(16N_\text{bin}\)-dimensional vector. 
This equation cannot be evaluated directly from the data because its right-hand side depends on the unknown parameter vector \(\mathbf x\) itself.

A standard method for solving the self-consistent equation  \(\mathbf{x}=T(\mathbf{x})\) is iteration.
Starting from an initial vector \(\mathbf x^{(0)}\), we successively evaluate
\(\mathbf x^{(1)}=T(\mathbf x^{(0)})\), \(\mathbf x^{(2)}=T(\mathbf x^{(1)})\), and, more generally, \(\mathbf x^{(n)}=T(\mathbf x^{(n-1)})\).
At the \(n\)-th iteration, \(\mathbf x^{(n-1)}\) is used to evaluate the discretized distribution \(f(\phi_\alpha)\) and the observables in Eq.~\eqref{eq:app-obs-tomo-with} for each event. The updated production-angle density is given by
\begin{equation}
   \mathcal R_\alpha^{(n)} = \frac{1}{N_{\rm event}}
\sum_{\rm events}
\widetilde O_{\mathcal R_\alpha}
(\Phi;\mathbf{x}^{(n-1)}), 
\end{equation}
while the updated polarization and spin-correlation parameters are given by the normalized ratios
\begin{equation}
    B_{i;\alpha}^{\pm(n)}
=
\frac{
\sum_{\rm events}
\widetilde O_{B^\pm_{i;\alpha}}
(\Phi;\mathbf{x}^{(n-1)})
}{
w_{\rm bin}
\sum_{\rm events}
\widetilde O_{\mathcal R_\alpha}
(\Phi;\mathbf{x}^{(n-1)})
},\quad C_{ij;\alpha}^{(n)}
= \frac{
\sum_{\rm events}
\widetilde O_{C_{ij;\alpha}}
(\Phi;\mathbf{x}^{(n-1)})
}{
w_{\rm bin}
\sum_{\rm events}
\widetilde O_{\mathcal R_\alpha}
(\Phi;\mathbf{x}^{(n-1)})
} \,.
\end{equation}
In the iterative reconstruction, the fold weights are evaluated using
the current trial parameters $\mathbf x^{(n-1)}$, while
$\mathcal R_\alpha^{(n)}$, $B_{i;\alpha}^{\pm,(n)}$, and
$C_{ij;\alpha}^{(n)}$ are evaluated together from the corresponding
fold-weighted sample averages. 
According to Eq.~\eqref{eq:app-self-consistent-equation}, these
estimators define the updated parameter vector $\mathbf x^{(n)}$.

\subsection{Local Convergence of the Fixed-point Iteration}
\label{sec:app-unfold-convergence}

In the infinite-statistics limit, the truth and its physical
displacements along the null direction form a continuous family of
fixed points,
\begin{equation}
\mathbf x_*(d_{\rm null})=
\mathbf x_{\rm truth}
+
d_{\rm null}\,\mathbf n_{\rm null},
\label{eq:app-null-fixed-point-family}
\end{equation}
where $\mathbf n_{\rm null}$ denotes the null deformation derived in
Sec.~\ref{sec:app-twofold-null}.  Each member of this family satisfies
\begin{equation}
T\!\left[\mathbf x_*(d_{\rm null})\right]=
\mathbf x_*(d_{\rm null}),
\label{eq:app-null-fixed-point-equation}
\end{equation}
provided that the corresponding density matrix remains physical.

We study the local convergence of the iteration toward this fixed-point
family.  We first introduce the gradient matrix of the iterative map,
$D_T(\mathbf x)$, with elements
\begin{equation}
(D_T)_{kl}\equiv\frac{\partial T_k}{\partial x_l},
\end{equation}
where $k,l=1,2,\ldots,16N_{\rm bin}$ label the components of the
parameter vector.  Differentiating
Eq.~\eqref{eq:app-null-fixed-point-equation} with respect to
$d_{\rm null}$ gives
\begin{equation}
D_T\!\left(\mathbf x_*(d_{\rm null})\right)\mathbf n_{\rm null}
=
\mathbf n_{\rm null}.
\label{eq:app-null-unit-eigenvalue}
\end{equation}
Thus, the null direction is an eigenvector of the gradient matrix with
eigenvalue one.  The iteration is therefore marginal along the
fixed-point family, and local attraction must be assessed in the
identifiable subspace transverse to the null direction.

If the spectral radius of $D_T(\mathbf x_*)$ restricted to the
identifiable subspace satisfies
\begin{equation}
R_{\rm id}\!\left(D_T\right)\big|_{\mathbf x_{*}}<1,
\label{eq:app-identifiable-spectral-radius}
\end{equation}
then perturbations in the identifiable directions are locally damped.
A smaller value of $R_{\rm id}$ generally corresponds to faster
asymptotic convergence toward the fixed-point family.

To verify this convergence structure numerically, we use the power
iteration method to determine the dominant and subdominant eigenmodes of the gradient matrix. 
The algorithm proceeds as follows:
\begin{itemize}
\item Choose a nonzero initial vector $\mathbf{v}_0$, typically taken to be random.
\item For $k=1,2,\ldots $, compute $\boldsymbol{\omega}_{k}=D_T \cdot \mathbf{v}_{k-1}, \mathbf{v}_{k}=\frac{\boldsymbol{\omega}_{k}}{||\boldsymbol{\omega}_{k}||}$, and estimate the eigenvalue at the $k$-th iteration by the Rayleigh quotient
$\lambda_{k}=\frac{\mathbf{v}_{k-1}^T \cdot \boldsymbol{\omega}_{k}}{\mathbf{v}_{k-1}^T \cdot\mathbf{v}_{k-1}}$.
\item Upon convergence, $\lambda_k$ approaches the eigenvalue with the largest modulus, while $\mathbf{v}_k$ approaches the corresponding eigenvector.
\end{itemize}

This method requires only matrix-vector products and does not require the full matrix to be constructed explicitly. It is therefore particularly suitable for the high-dimensional $16N_{\rm bin}\times16N_{\rm bin}$ gradient matrix considered here.
In practice, this matrix-vector product is evaluated numerically using a central finite-difference method:
\begin{equation}
    D_T(\mathbf{x}_*) \cdot \mathbf{v} \simeq\frac{T(\mathbf{x}_*+\epsilon \mathbf{v})-T(\mathbf{x}_*-\epsilon \mathbf{v})}{2\epsilon}
    \label{eq:central-finite-difference}
\end{equation}
where $\epsilon$ is a small real step size.

\subsubsection{Standard Model Benchmark}
\label{sec:app-unfold-convergence-SM}

We first apply the power iteration method to the Standard Model sample. 
In this test, the fixed point \(\mathbf{x}_*\) is chosen to be the theoretical Standard Model value of the binned parameter vector, rather than the output of the self-consistent reconstruction. The action of \(D_T(\mathbf{x}_*)\) on an arbitrary vector is then evaluated by the central finite-difference method described above. Here we use \(\epsilon=10^{-3}.\)

In the numerical implementation, we generate 100 statistically independent Monte Carlo samples with MadGraph5\_aMC@NLO \cite{Alwall:2014hca}, each containing \(10^6\) events.
The power iteration is performed independently on each sample for 140 steps.
At the final step, the difference between $\lambda_k$ and $||\boldsymbol{\omega}_k||$ is smaller than $10^{-6}$, and the eigenvector residual $\frac{||D_T \cdot \mathbf{v}-\lambda \mathbf{v}||}{||D_T \cdot \mathbf{v}||}$ is also below $10^{-6}$. These checks indicate that the power iteration has converged with sufficient numerical accuracy.

Here, the dominant eigenvalue refers to the eigenvalue with the largest modulus, and the dominant eigenvector is the corresponding eigenvector. The dominant eigenvalue is then averaged over the 100 independent samples, yielding
\begin{equation}
    \lambda_{\rm dom}= 1.00097 \pm 0.00025,
    \label{eq:SM-dominant-eigenvalue}
\end{equation}
where the uncertainty represents the standard error of the mean over the 100 independent samples.

\begin{figure}[htb]
    \centering
    \includegraphics[width=0.6\linewidth]{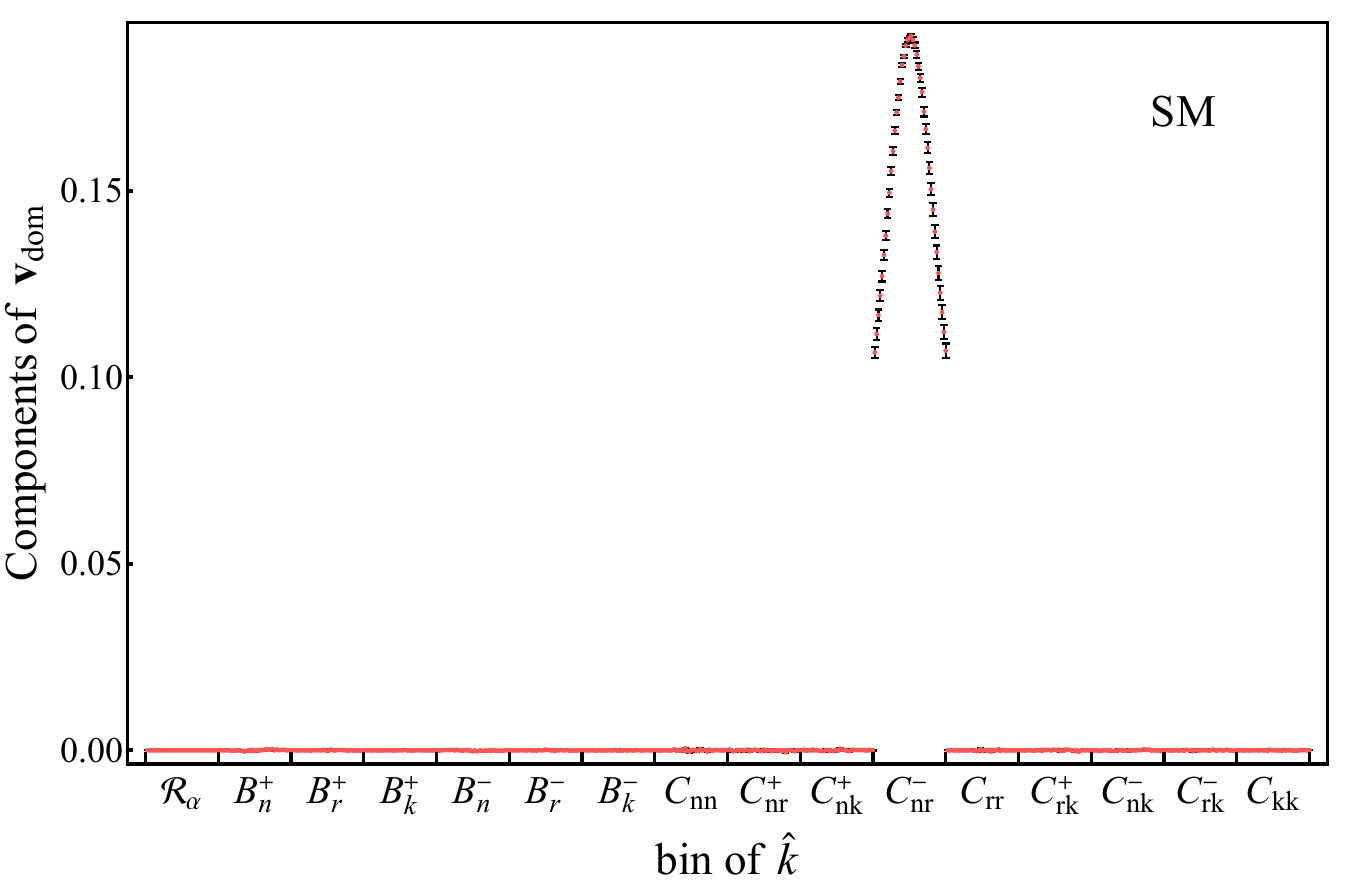}
    \caption{
The components of dominant eigenvector \(\mathbf{v}_{\rm dom}\) for the Standard Model sample. The horizontal axis denotes the components of the flattened binned parameter vector: the 16 parameter groups are shown in sequence, and each parameter group contains 40 bins in \(\cos\theta\). The vertical axis gives the corresponding component of the normalized eigenvector. The eigenvector is localized almost entirely in the antisymmetric component \(C^{-}_{nr}\), or equivalently \(C_{nr}-C_{rn}\), while all other components are consistent with zero. This agrees with the analytic result that \(C_{nr}-C_{rn}\) is the only unidentifiable direction in the tau-pair pion channel.
}
    \label{fig:SM-dominant-eigenvector}
\end{figure}

The corresponding normalized dominant eigenvector $\mathbf{v}_{\rm dom}$ of \(D_T(\mathbf{x}_*)\), obtained from the power iteration, is shown in Fig.~\ref{fig:SM-dominant-eigenvector}.
The horizontal axis denotes the components of the flattened binned parameter vector. These 16 groups correspond to the normalized production-rate parameter \(\mathcal{R}_\alpha\), the polarization components \(B^{+}_{i,\alpha}\) and \(B^{-}_{i,\alpha}\), and the spin-correlation components \(C_{ij,\alpha}\).
For the off-diagonal spin correlations, we use the combinations
\(
C^{+}_{ij,\alpha}\equiv C_{ij,\alpha}+C_{ji,\alpha},
\;
C^{-}_{ij,\alpha}\equiv C_{ij,\alpha}-C_{ji,\alpha},
\)
which represent the symmetric and antisymmetric components, respectively. Within each group, the 40 entries correspond to the 40 production-angle bins, or equivalently, the 40 bins in \(\cos\theta\). 
The vertical axis gives the value of the normalized eigenvector in each component of this flattened parameter space.

The dominant eigenvector $\mathbf{v}_{\rm dom}$ is almost entirely localized in the antisymmetric spin-correlation component \(C^{-}_{nr}\), corresponding to the combination \(C_{nr}-C_{rn}\). 
All other components, including the production-rate parameter, the polarization vectors, and the remaining spin-correlation components, are consistent with zero within the statistical fluctuations across the independent Monte Carlo samples.

This result has a direct interpretation. The dominant eigenvalue is statistically close to unity, as expected for a null direction, with the small upward shift arising from finite-sample and numerical fluctuations. So the full iterative map is not locally convergent when this direction is included. 
However, the non-convergent mode is confined to the \(C_{nr}-C_{rn}\) direction. 
This is precisely the component identified by the informational-completeness analysis as the null direction of the twofold coarse-grained measurement. 
Therefore, the power iteration results show that the apparent instability of the iteration is not a failure of the reconstruction of the physically identifiable components. 
Instead, it reflects the fact that the visible pion distribution contains no information about the antisymmetric combination \(C_{nr}-C_{rn}\). Once this null direction is removed or fixed by an external convention, the remaining identifiable components can be reconstructed by the iterative method.

To verify that the non-convergent behavior is confined to the null direction, we further perform a deflated power iteration to compute the subdominant eigenvalue of \(D_T(\mathbf{x}_*)\).
After obtaining the dominant eigenvector \(v_{\rm dom}\), we subtract the component along \(v_{\rm dom}\) at each step. This suppresses the mode associated with \(C_{nr}-C_{rn}\) and allows the iteration to probe the remaining parameter directions.

We explicitly check the result by computing the residual with respect to the original,
\(
\frac{||D_T(\mathbf{x}_*) \cdot \mathbf{v}_{\rm sub}-\lambda_{\rm sub} \mathbf{v}_{\rm sub}||}
{||D_T(\mathbf{x}_*)\cdot \mathbf{v}_{\rm sub}||}
\), where $\mathbf{v}_{\rm sub}$ and $\lambda_{\rm sub}$ are the subdominant  eigenvector and the subdominant (second largest) eigenvalue.
The residual is found to be small (about \(10^{-3}\)), indicating that the deflated iteration has converged to an eigenmode of the original gradient matrix to good numerical accuracy. 
The convergent subdominant eigenvalue is
\begin{equation}
    \lambda_{\rm sub}=0.96046\pm 0.00012 ,
    \label{eq:SM-subdominant-eigenvalue}
\end{equation}
which is clearly smaller than unity.
Therefore, once the non-identifiable \(C_{nr}-C_{rn}\) direction is removed, all remaining perturbation modes are damped by the iterative map.

Since $D_T(\mathbf{x}_*)$ is not guaranteed to be symmetric or
normal, we also perform a matrix-free Arnoldi analysis~\cite{Saad1980}
as an independent cross-check, using the same finite-difference
Jacobian-vector products as in Eq.~\eqref{eq:central-finite-difference}. We compute the ten
largest-magnitude Ritz values, namely, the eigenvalue approximations obtained by projecting $D_T(\mathbf{x}_*)$ onto the Krylov subspace
constructed in the Arnoldi iteration. The two largest-magnitude Ritz values are
$1.00121\pm0.00015$ and $0.96043\pm0.00012$, in good agreement with the power-iteration results in Eqs.~\eqref{eq:SM-dominant-eigenvalue} and \eqref{eq:SM-subdominant-eigenvalue}, while the remaining eight
computed Ritz values have smaller moduli. For each of the ten Ritz
pairs, the mean normalized eigenpair residual
$\|D_T(\mathbf{x}_*)\mathbf{v}-\lambda\mathbf{v}\|/
\|D_T(\mathbf{x}_*)\mathbf{v}\|$ over the 100 independent samples is
below $5.7\times10^{-7}$, and the largest residual observed over all
modes and samples is $4.1\times10^{-6}$.
This confirms that the iterative reconstruction is locally convergent on the identifiable parameter subspace and that the only non-convergent mode is the null direction identified by the informational-completeness analysis.

\subsubsection{Anomalous Tau-dipole Benchmark}
\label{sec:app-unfold-convergence-dipole}

We next apply the same power-iteration analysis to a non-Standard-Model benchmark. 
This test checks whether the numerical convergence analysis depends on the Standard Model spin structure. 
We consider anomalous magnetic- and electric-dipole interactions of the tau lepton.
The anomalous magnetic moment is denoted by \(a_\tau\), which is dimensionless, while the electric dipole moment \(d_\tau\) is rewritten in terms of the dimensionless parameter
\(
\tilde d_\tau \equiv \frac{2m_\tau}{e}d_\tau .
\)
In this benchmark, both couplings are allowed to be complex, and we choose
\begin{equation}
a_\tau = 0.01 - 0.02i,
\qquad
\tilde d_\tau = -0.05 + 0.03i.
\end{equation}
Monte Carlo events are generated with these anomalous couplings, and both the power-iteration and the deflated power-iteration are performed in the same way as in the Standard Model case. 
The fixed point \(\mathbf{x}_*\) used in the finite-difference evaluation of \(D_T(\mathbf{x}_*)\) is taken to be the theoretical prediction for the binned parameter vector at this benchmark point.

\begin{figure}[htb]
    \centering
    \includegraphics[width=0.6\linewidth]{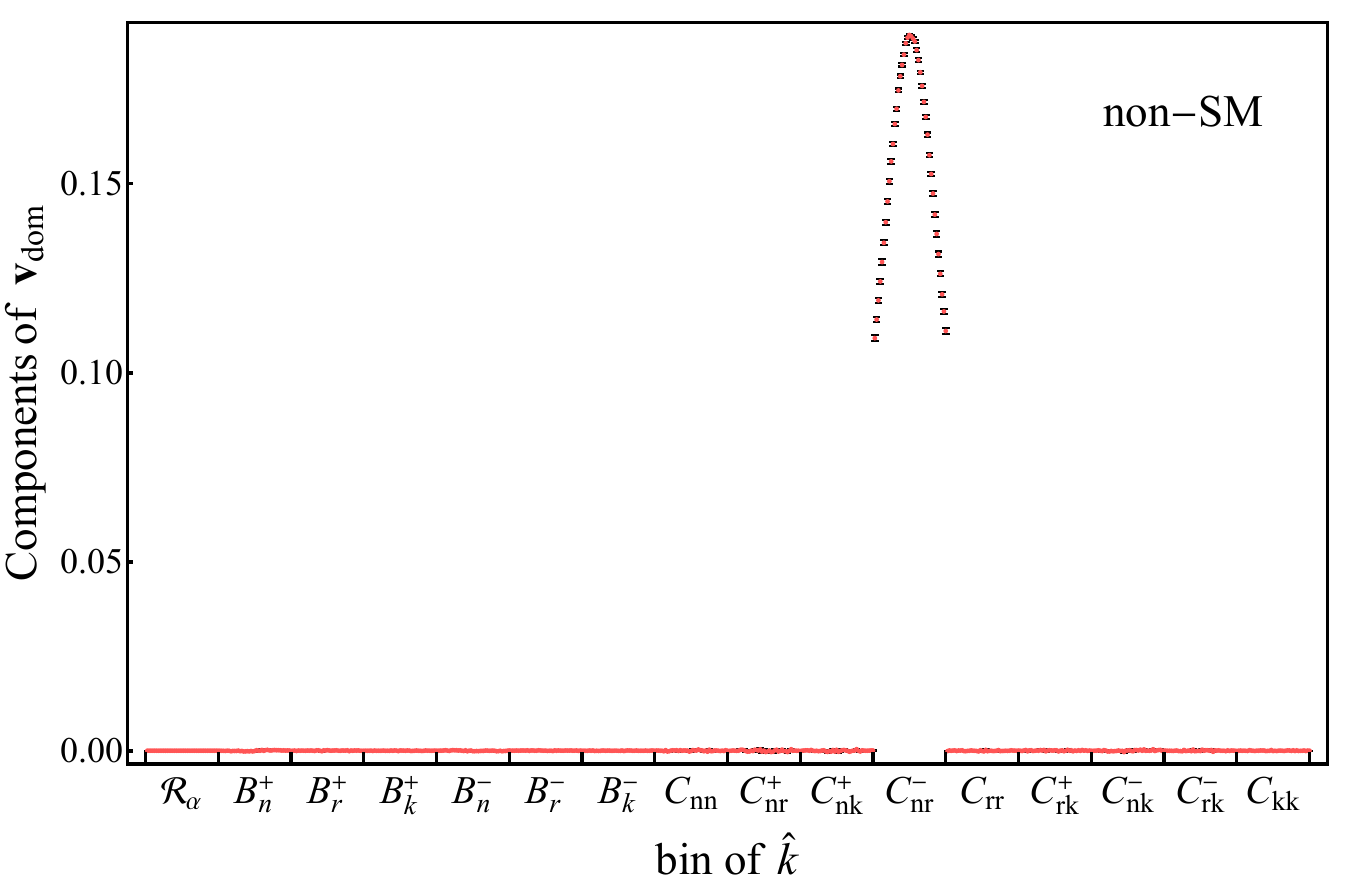}
    \caption{
The components of dominant eigenvector $\mathbf{v}_{\rm dom}$ of \(D_T(\mathbf{x}_*)\) for the anomalous tau-dipole benchmark with \(a_\tau=0.01-0.02i\) and \(\tilde d_\tau=-0.05+0.03i\). The fixed point \(\mathbf{x}_*\) is taken to be the theoretical prediction for this benchmark. The eigenvector is again localized in \(C^{-}_{nr}\), corresponding to \(C_{nr}-C_{rn}\), while the other components are consistent with zero. 
This confirms that the least-damped mode is the same null direction as in the Standard Model case.
}

\label{fig:BSM-dominant-eigenvector}
\end{figure}

The dominant and subdominant eigenvalues are
\begin{equation}
\lambda_{\rm dom}=1.00128\pm 0.00012,\qquad \lambda_{\rm sub} = 0.96082\pm0.00013 .
\label{eq:BSM-eigenvalues}
\end{equation}
The corresponding dominant eigenvector is shown in Fig.~\ref{fig:BSM-dominant-eigenvector}.
As in the Standard Model case, the dominant eigenvector is localized almost entirely in the antisymmetric spin-correlation component \(C^{-}_{nr}\), or equivalently \(C_{nr}-C_{rn}\), while the other components are consistent with zero within statistical fluctuations.
The subdominant eigenvalue is also smaller than unity, as in the Standard Model case. 
This shows that the least-damped direction is again the null direction identified by the informational-completeness analysis, even though the underlying spin density matrix is deformed by the anomalous dipole interactions.

As an independent cross-check, we repeat the same matrix-free Arnoldi analysis for the anomalous tau-dipole benchmark. The two largest-magnitude Ritz values are
$1.00130\pm0.00012$ and $0.96079\pm0.00012$, in good agreement with
the dominant and subdominant eigenvalues in
Eq.~\eqref{eq:BSM-eigenvalues}, while the remaining eight computed
Ritz values have smaller moduli. For each of the ten Ritz pairs, the mean normalized eigenpair residual over the 100 independent samples
is below $5.6\times10^{-7}$, and the largest residual observed over all modes and samples is $4.1\times10^{-6}$.

It should be mentioned that the benchmark values used here are deliberately chosen to be very large, well beyond the range expected for realistic tau dipole moments, so that they provide a stress test of the numerical procedure rather than a phenomenologically realistic benchmark.
In particular, we do not observe a level crossing between the dominant and subdominant modes, nor a qualitative change in their stability properties.
This indicates that the convergence structure of the iterative map is stable under sizable deformations of the underlying spin density matrix.
Hence, the method is robust in the presence of possible new-physics contributions.

\section{Iterative Tomography Results for Tau Pairs}
\label{sec:app-result}

\subsection{Standard Model Benchmark}
\label{sec:app-result-SM}

We now present the iterative reconstruction results for the full set of binned production-density-matrix parameters in the Standard Model sample. These include the normalized production-angle distribution $\mathcal R_\alpha$, the polarization components $B^+_{i,\alpha}$ and $B^-_{i,\alpha}$, and the spin-correlation components $C_{ij,\alpha}$. Altogether, the reconstruction contains 16 binned parameter functions, each shown as a function of the production angle.

The iteration is initialized with a spin-independent trial density matrix: all polarization and spin-correlation components are set to zero, and the production-angle distribution is taken to be uniform in $\cos\theta$. Since the two kinematic solutions have equal Jacobians in this channel, this initial choice assigns equal weights to the two solutions for every event,
$\omega_a=\omega_b=1/2$,
in the first update. The first iteration is therefore equivalent to the flat $50/50$ average over the twofold ambiguity and provides the baseline for comparison with the self-consistent reconstruction.

\begin{figure}[!t]
  \centering
  \begin{minipage}{0.49\textwidth}
    \centering
    \includegraphics[width=\linewidth]{figures/SM_iteration_4panel_with_legend.pdf}
    \label{fig:sm_iteration_1}
  \end{minipage}
  \hfill
  \begin{minipage}{0.49\textwidth}
    \centering
    \includegraphics[width=\linewidth]{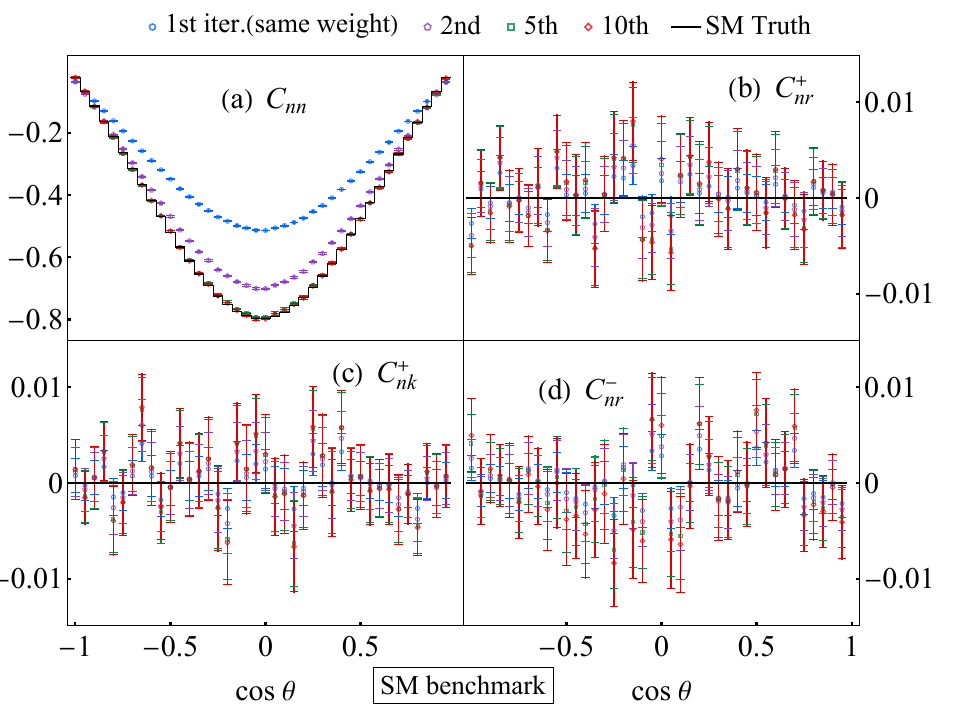}
    \label{fig:sm_iteration_2}
  \end{minipage}
\\
  \centering
  \begin{minipage}{0.49\textwidth}
    \centering
    \includegraphics[width=\linewidth]{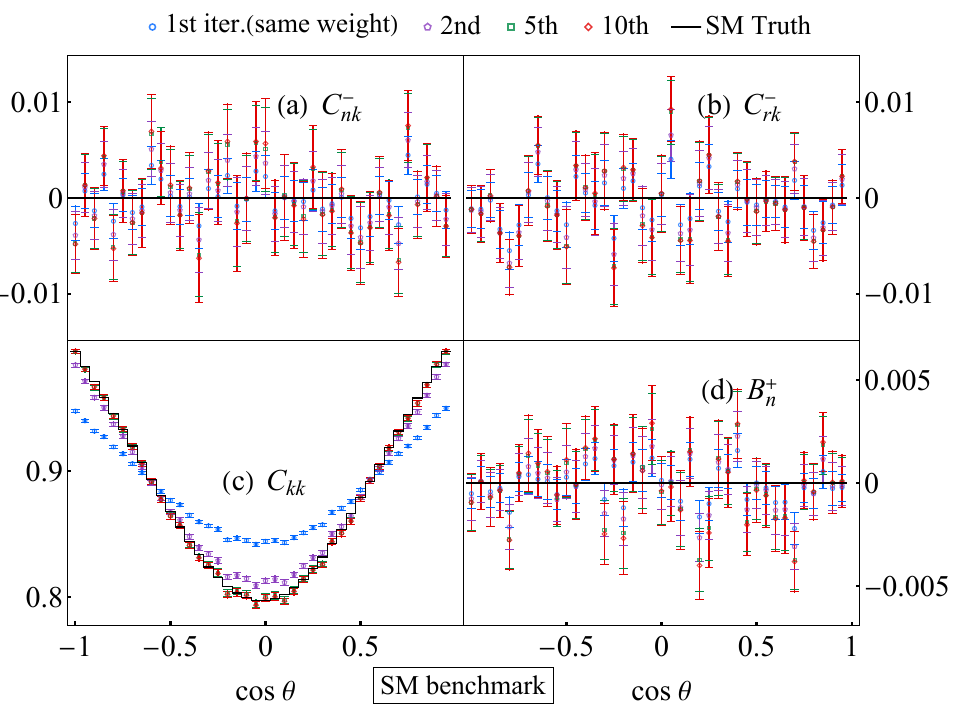}
    \label{fig:sm_iteration_3}
  \end{minipage}
  \hfill
  \begin{minipage}{0.49\textwidth}
    \centering
    \includegraphics[width=\linewidth]{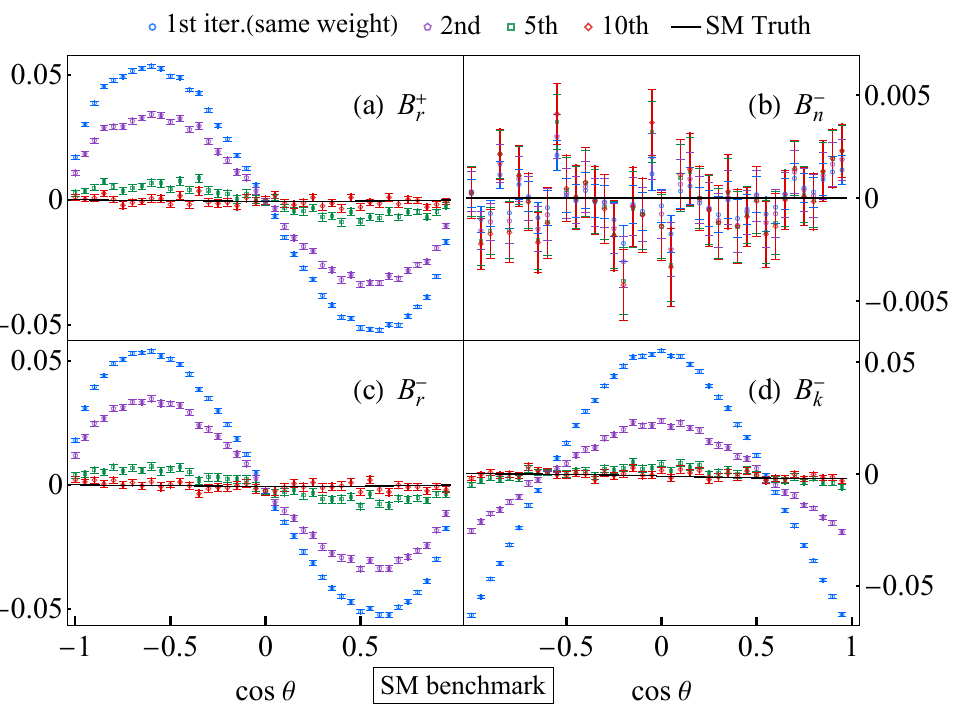}
    \label{fig:sm_iteration_4}
  \end{minipage}
  \caption{
Iterative reconstruction of the full set of 16 binned production-density-matrix parameters in the Standard Model
$e^+e^-\to\tau^+\tau^-\to\pi^+\pi^-+\nu\bar\nu$
sample. The truth values are compared with the first, second, fifth, and tenth iterations. The first iteration corresponds to the flat $50/50$ average over the two kinematic solutions. The self-consistent iterations correct the flat-average biases and approach the truth values for all identifiable components. 
}
\label{fig:sm-iteration-all}
\end{figure}

Figure~\ref{fig:sm-iteration-all} compares the first, second, fifth, and tenth iterations with the Standard Model truth values for all 16 parameter functions. The first iteration exhibits visible biases in several components. After successive self-consistent updates, the reconstructed identifiable parameters move toward the truth values. For these components, the fifth and tenth iterations are already close to each other, indicating that the reconstruction has largely stabilized by the tenth iteration. The identifiable components agree well with the truth values, demonstrating that the self-consistent fold weights correct the biases introduced by the flat-average prescription.

For the SM benchmark, several components, including $B_n^\pm$, $C_{nr}^+$, $C_{nk}^\pm$, and $C_{rk}^-$, have vanishing or numerically small truth values and therefore appear close to zero on the scale of the corresponding panels. Their reconstructed values fluctuate around these small truth values because of finite-sample statistical uncertainties. When the flat $50/50$ average already gives a result close to the truth, the subsequent iterations need not display a visually smooth convergence pattern; instead, they remain statistically distributed around the small or vanishing truth value.

For other components with vanishing or numerically small SM truth values, such as $B_r^\pm$ and $B_k^\pm$, the flat prescription produces a visible spurious offset, which is progressively removed by the self-consistent updates. After the bias is removed, the final iterations still fluctuate around the small or vanishing truth value, because of finite-sample statistical uncertainties.

The null component $C^-_{nr}$ should be interpreted differently.
In the SM benchmark, the spin-independent initialization sets all
spin-correlation components, including $C^-_{nr}$, to zero. Since the SM truth value of this null component is also zero, the flat first iteration and the subsequent iterated values remain close to the truth in this direction, up to finite-sample fluctuations. This behavior, however, should not be interpreted as an independent reconstruction of $C^-_{nr}$ from the visible data.

As shown by the analytic null-space analysis in Sec.~\ref{sec:app-twofold-null}, variations along $C^-_{nr}$ belong to the kernel of the twofold visible-data map. Equivalently, within this twofold pion-channel measurement, the visible distribution does not provide an independent constraint on this antisymmetric spin-correlation component. Its numerical value in the
iteration is therefore controlled by the initialization and finite-sample fluctuations, rather than being fixed by the visible data. This point is made more explicit in the non-SM benchmark below, where the truth value
of $C^-_{nr}$ is nonzero and the iteration does not approach it.

The overall convergence is quantified by the relative RMS residual defined in Eq.~\eqref{eq:global-residual} of the main text. This residual is evaluated over all identifiable binned parameters, with the null combination $C_{nr}^-=C_{nr}-C_{rn}$ excluded. As shown in Fig.~\ref{fig:tau-residuals} of the main text, it decreases rapidly with iteration.

\subsection{Anomalous Tau-dipole Benchmark}
\label{sec:app-result-BSM}

\begin{figure}[htb]
  \centering
  \begin{minipage}{0.49\textwidth}
    \centering
    \includegraphics[width=\linewidth]{figures/BSM_iteration_4panel_with_legend.pdf}
    \label{fig:bsm_iteration_1}
  \end{minipage}
  \hfill
  \begin{minipage}{0.49\textwidth}
    \centering
    \includegraphics[width=\linewidth]{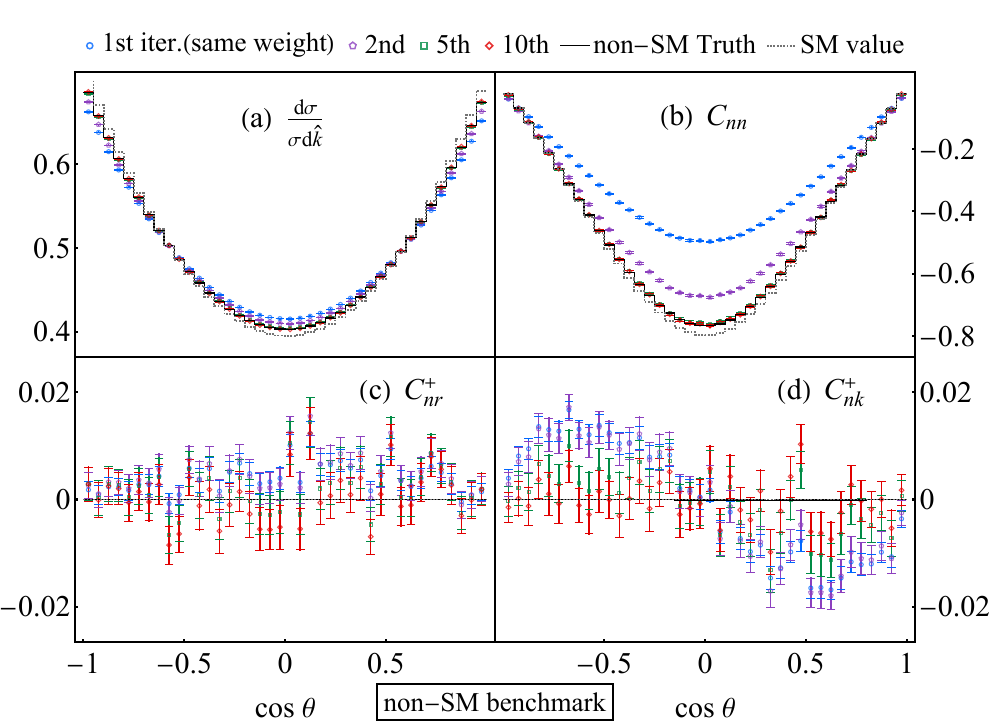}
    \label{fig:bsm_iteration_2}
  \end{minipage}
  \begin{minipage}{0.49\textwidth}
    \centering
    \includegraphics[width=\linewidth]{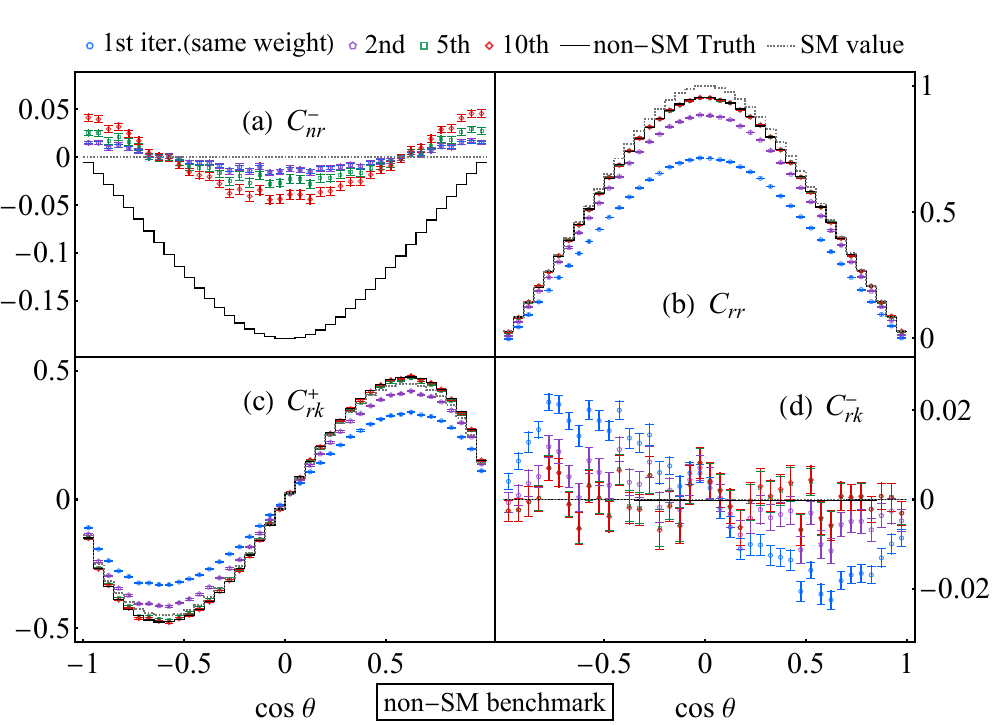}
    \label{fig:bsm_iteration_3}
  \end{minipage}
  \hfill
  \begin{minipage}{0.49\textwidth}
    \centering
    \includegraphics[width=\linewidth]{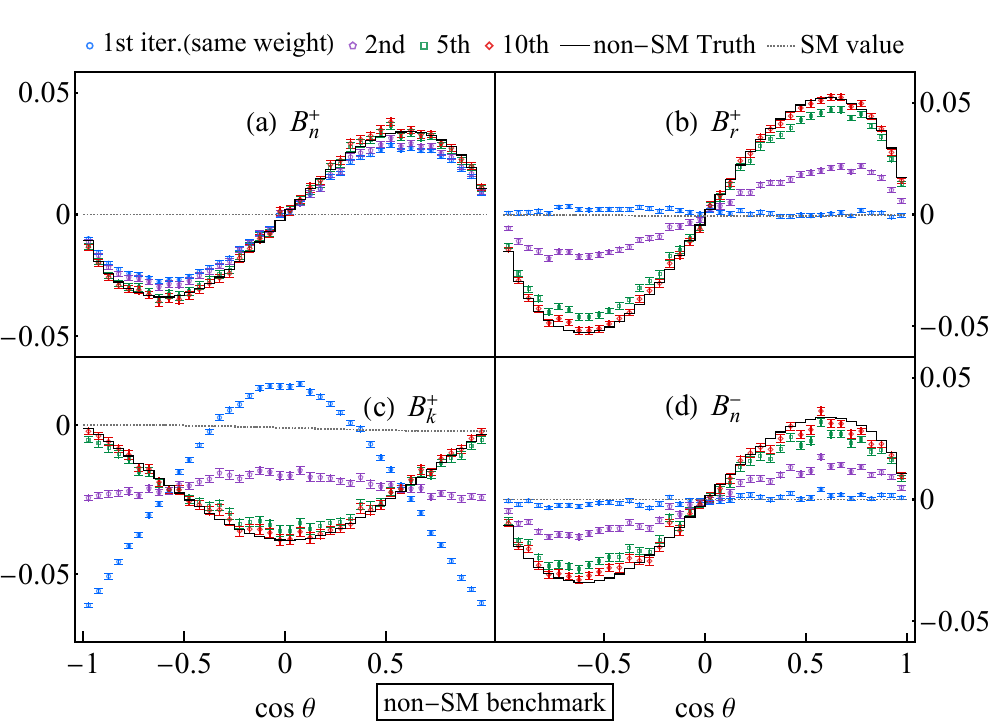}
    \label{fig:bsm_iteration_4}
  \end{minipage}
\caption{
Iterative reconstruction of the full set of 16 binned
production-density-matrix parameters in the non-Standard Model
$e^+e^-\to\tau^+\tau^-\to\pi^+\pi^-+\nu\bar\nu$ sample with anomalous
tau-dipole interactions. The non-SM truth values are compared with the
first, second, fifth, and tenth iterations, while the corresponding SM
truth values are also shown for reference. The first iteration
corresponds to the flat $50/50$ average over the two kinematic
solutions. The self-consistent iterations correct the flat-average
biases and approach the non-SM truth values for all identifiable
components. The null component $C_{nr}^-$ exhibits qualitatively
different behavior, as examined in detail below.
}
\label{fig:BSM-numeric-reconstruction}
\end{figure}

We next present in Fig.~\ref{fig:BSM-numeric-reconstruction} the
iterative reconstruction for the anomalous tau-dipole benchmark
introduced in Sec.~\ref{sec:app-unfold-convergence-dipole}. The non-SM
truth values differ visibly from the corresponding SM predictions,
which are also shown for comparison. Nevertheless, the self-consistent
iterations rapidly approach the non-SM truth values for all
identifiable components, confirming that the reconstruction does not
rely on an SM production template.

The null component $C^-_{nr}$ exhibits qualitatively different
iteration behavior. It also approaches a stable numerical plateau
after sufficiently many iterations, but this plateau does not in
general coincide with the nonzero non-SM truth value. Instead, the
difference between the iterated result and the truth approaches the
null deformation predicted in Eq.~\eqref{eq:tau-null-short} of the main
text:
\begin{equation}
\delta C_{nr}(\hat k)
=-\delta C_{rn}(\hat k)=
d_{\rm null} \times
\left(\frac{\mathrm d\sigma}{\sigma\,\mathrm d\hat{k}}\right)^{-1} .
\nonumber
\end{equation}

More explicitly, we have defined
$
R_\alpha
\equiv
\left.
\frac{\mathrm d\sigma}
{\sigma \,\mathrm d\hat{k}}
\right|_\alpha $ to
denote the normalized production-angle distribution in bin $\alpha$.
For a result obtained after sufficiently many iterations, we define
\begin{equation}
d_{{\rm null},\alpha}
\equiv
\frac{1}{2}R_\alpha
\left[
\left(C^-_{nr,\alpha}\right)_{\rm iter} -
\left(C^-_{nr,\alpha}\right)_{\rm truth}
\right].
\label{eq:dnull-bin-definition}
\end{equation}
The factor of $1/2$ follows from
$C^-_{nr}=C_{nr}-C_{rn}$ together with
$\delta C_{nr}=-\delta C_{rn}$ in
Eq.~\eqref{eq:tau-null-short}. If the iterated result differs from the
truth only along the analytic null direction, then
$d_{{\rm null},\alpha}$ must be independent of the production-angle
bin $\alpha$:
\begin{equation}
d_{{\rm null},\alpha}=d_{\rm null}.
\end{equation}
Equivalently, the difference between the iterated density matrix and
the truth belongs to the one-dimensional null space of the twofold
visible-data map.

\begin{figure}[htb]
    {
\centering
\begin{minipage}[t]{0.49\textwidth}
    \centering
    \includegraphics[width=\linewidth]{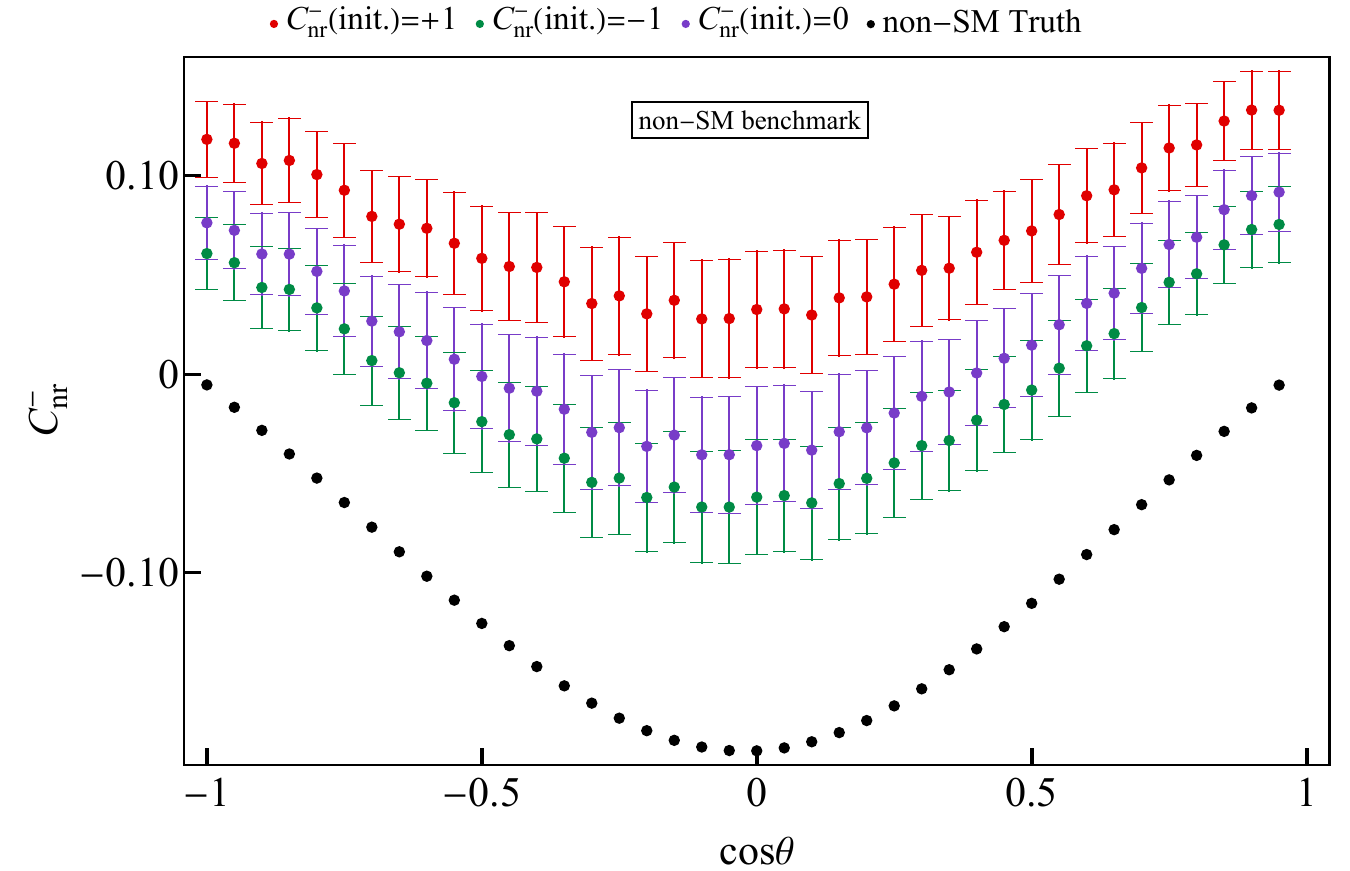}
    
    \textbf{(a)}
\end{minipage}
\hfill
\begin{minipage}[t]{0.49\textwidth}
    \centering
    \includegraphics[width=\linewidth]{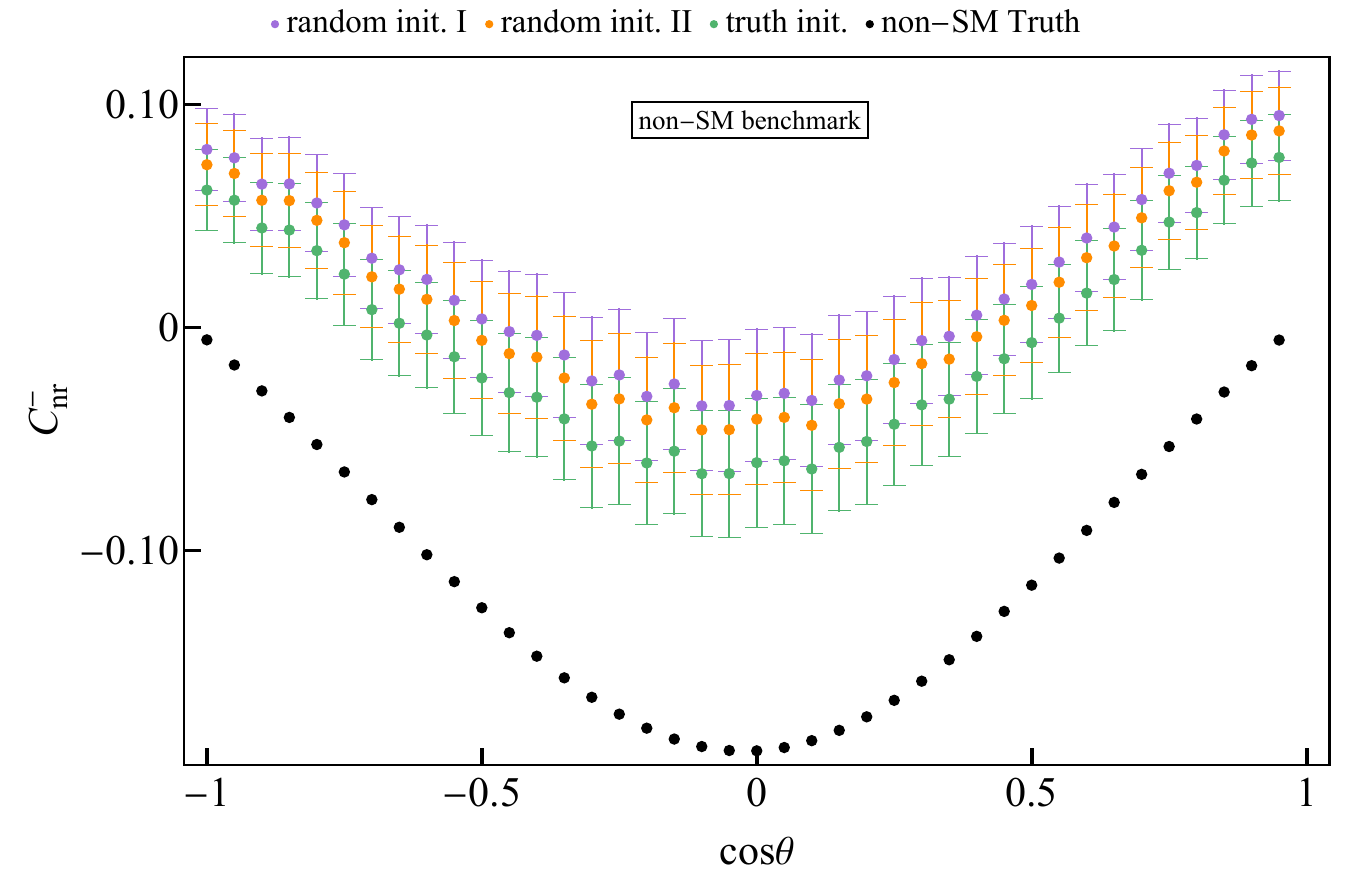}
    
    \textbf{(b)}
\end{minipage}
\par
}
\caption{
Long-iteration behavior of the null component $C^-_{nr}$ in the
anomalous tau-dipole benchmark for different initializations.
Panel (a) shows three simple initializations, for which the
production-angle distribution is initialized to be uniform and all
polarization and spin-correlation components are initialized to zero,
except that $C^-_{nr}$ is initialized to $0$, $+1$, and $-1$,
respectively. Panel (b) shows two random initializations and one
initialization at the theoretical truth value. For the random
initializations, the production-angle distribution is randomly
perturbed with an amplitude between $0.10$ and $0.20$, the polarization
components satisfy $|B_i^+|,|B_i^-|\leq0.20$, and the spin-correlation
matrix is rescaled to have Frobenius norm no larger than $0.30$.
All runs are evolved for 4000 iterations. The plateau value of
$C^-_{nr}$ depends on the initialization, but its shift from the truth
is consistent with the null deformation in
Eq.~\eqref{eq:tau-null-short}. In particular,
$d_{{\rm null},\alpha}=
R_\alpha[(C^-_{nr,\alpha})_{\rm iter}
-(C^-_{nr,\alpha})_{\rm truth}]/2$
is approximately independent of $\alpha$, up to finite-sample
fluctuations.
}
\label{fig:cnrminus-initialization-dependence}
\end{figure}

To test this fixed-point degeneracy numerically, we repeat the non-SM
iteration with six different initial conditions and evolve each run
for 4000 iterations. The first three initializations use a uniform
production-angle distribution and set all polarization and
spin-correlation components to zero, except that the initial value of
$C^-_{nr}$ is chosen to be $0$, $+1$, and $-1$, respectively. The
remaining three cases consist of two random initializations and one
initialization at the theoretical truth value. For the random
initializations, the production-angle distribution is randomly
perturbed with an amplitude between $0.10$ and $0.20$, the polarization
components satisfy
$|B_i^+|,|B_i^-|\leq 0.20$, and the spin-correlation matrix is rescaled
to have Frobenius norm no larger than $0.30$. This condition also
ensures that its spectral norm is no larger than $0.30$.

For all six initial conditions, the identifiable components exhibit
essentially the same iteration behavior and approach the non-SM truth
values within approximately ten iterations. By contrast, the null
component $C^-_{nr}$ evolves on a much longer iteration scale. In our
tests, it typically requires between 2000 and 4000 iterations to reach
an approximately stable plateau. The results after 4000 iterations are
shown in Fig.~\ref{fig:cnrminus-initialization-dependence}.

The different initializations lead to different plateau values of
$C^-_{nr}$. In every case, however, the corresponding
$d_{{\rm null},\alpha}$ is approximately constant across the
production-angle bins. We extract a single null-deformation parameter
for each initialization by defining
\begin{equation}
\bar d_{\rm null} =
\frac{1}{N_{\rm bin}}
\sum_\alpha d_{{\rm null},\alpha}.
\label{eq:dnull-bin-average}
\end{equation}
The resulting values are listed in
Table~\ref{tab:dnull-initializations}. The small bin-to-bin spreads
show that the differences from the truth are not arbitrary
reconstruction biases, but follow the specific production-angle
dependence predicted analytically for the null direction. The
dependence of $\bar d_{\rm null}$ on the initialization further
demonstrates that this coordinate is not selected uniquely by the
visible data.

The truth-initialized run provides a particularly instructive example.
Although all components are initially set to their theoretical truth
values, the identifiable components remain close to, or rapidly return
to, their truth values, whereas $C^-_{nr}$ gradually moves away from
the truth and reaches a nonzero value of $\bar d_{\rm null}$. This does
not contradict the infinite statistics level fixed-point relation
$T(\mathbf x_{\rm truth})=\mathbf x_{\rm truth}$. 
The numerical iteration is constructed from a finite Monte Carlo sample
and therefore uses an empirical update map $\widehat T_N$, rather than
the exact infinite-statistics map $T$. Consequently, for a particular
finite sample, the theoretical truth need not be left exactly invariant:
\begin{equation}
\widehat T_N(\mathbf x_{\rm truth})
\neq
\mathbf x_{\rm truth}.
\end{equation}

\begin{table}[t]
\caption{
Extracted values of the null-deformation parameter
$\bar d_{\rm null}$ for the anomalous tau-dipole benchmark after 4000
iterations with different initializations. The second column gives the
bin-averaged value $\bar d_{\rm null}$, where the uncertainty is
obtained by propagating the statistical uncertainties of the binned
quantities. The third column gives the bin-to-bin standard deviation
of the central values of $d_{{\rm null},\alpha}$ and therefore
quantifies the consistency of the observed deformation with a
bin-independent constant.
}
\label{tab:dnull-initializations}
\centering
\begin{ruledtabular}
\begin{tabular}{lcc}
Initialization & $\bar d_{\rm null}$ & bin-to-bin spread \\
\hline
zero initialization & $0.0308\pm0.0010$ & 0.00161 \\
$C^-_{nr}=+1$ initialization & $0.0448\pm0.0010$ & 0.00151 \\
$C^-_{nr}=-1$ initialization & $0.0254\pm0.0009$ & 0.00157 \\
random seed 1 & $0.0320\pm0.0010$ & 0.00159 \\
random seed 2 & $0.0297\pm0.0010$ & 0.00159 \\
truth initialization & $0.0258\pm0.0010$ & 0.00156 \\
\end{tabular}
\end{ruledtabular}
\end{table}

Statistical fluctuations generated by
the first few updates are damped in the identifiable subspace, but
there is no data-driven restoring force that returns a displacement
along the null direction to $d_{\rm null}=0$. Consequently, even exact
truth initialization does not protect the unidentifiable component
against finite-sample drift.

Consistent with the analytic unit-eigenvalue relation in
Eq.~\eqref{eq:app-null-unit-eigenvalue}, the power-iteration analysis
in Sec.~\ref{sec:app-unfold-convergence} finds that the dominant
numerical mode is strongly localized in $C^-_{nr}$ and has an
eigenvalue close to unity.
In the finite-sample calculation, the estimated eigenvalue can
lie slightly above unity near the truth input. This small deviation
should be interpreted as a finite-sample and numerical lifting of the
marginal null mode of the exact infinite-statistics map, rather than as
a modification of the analytic null-space result. It accounts for the
slow departure from the truth along $C^-_{nr}$, while the subsequent
nonlinear evolution determines the initialization- and
sample-dependent plateau. The fact that every observed displacement in Fig.~\ref{fig:cnrminus-initialization-dependence}
follows Eq.~\eqref{eq:tau-null-short} provides a direct numerical
verification of the analytic null space and of the associated
continuous fixed-point degeneracy.

\section{Positivity-constrained Concurrence and CHSH Ranges}
\label{sec:app-concurrence}

Concurrence is an entanglement measure for a pair of spin-$1/2$ particles.
Its value ranges from 0 to 1, where $\mathcal{C} = 0$ indicates a completely separable state (no entanglement) and $\mathcal{C} = 1$ corresponds to a maximally entangled state.

For a general mixed state with non-zero polarization parameters, the Concurrence cannot be written as a simple algebraic formula depending only on $B_i$ and $C_{ij}$. 
Instead, the full $4 \times 4$ density matrix $\rho(B_i^+, B_j^-, C_{ij})$ must first be constructed.
The spin-flipped density matrix is defined as
$\tilde{\rho}=(\sigma_y\otimes\sigma_y)\rho^*(\sigma_y\otimes\sigma_y)$,
where $\sigma_y$ is the Pauli $y$ matrix and $\rho^*$ denotes the complex conjugate of $\rho$.
Let $\lambda_1 \ge \lambda_2 \ge \lambda_3 \ge \lambda_4$ be the eigenvalues of the Hermitian matrix $\sqrt{\sqrt{\rho}\tilde{\rho}\sqrt{\rho}}$. The concurrence is then defined by the formula:
\begin{equation}
\mathcal{C}(\rho) = \max(0, \lambda_1 - \lambda_2 - \lambda_3 - \lambda_4).
\end{equation}

The CHSH parameter is a physical observable used to test the Clauser-Horne-Shimony-Holt (CHSH) Bell inequality. It serves to distinguish whether a system can be explained by local hidden variable theories, or if it exhibits non-local quantum behavior.
In any classical framework obeying local realism, the absolute value of this parameter is bounded by $|\mathcal{B}| \le 2$. 
In quantum mechanics, strong entanglement allows this inequality to be violated, reaching a maximum theoretical value of $2\sqrt{2}$.

Consider two separated observers performing joint measurements on a pair of particles. 
The first observer can choose a measurement operator of $A$ or $A'$ on the first particle, while the second observer can choose a measurement operator of $B$ or $B'$ on the second particle. 
Then, the CHSH parameter is defined as a linear combination of the expectation values of these joint measurements:
\begin{equation}
\mathcal{B} = \langle A \otimes B \rangle + \langle A \otimes B' \rangle + \langle A' \otimes B \rangle - \langle A' \otimes B' \rangle.
\end{equation}
The maximal CHSH value $\mathcal{B}_{\max}$, optimized over the measurement directions, is determined entirely by the spin-correlation matrix $C_{ij}$ and is independent of the polarization vectors $B_i^\pm$. Defining the real symmetric matrix $M\equiv C^T C$, one obtains the analytical expression of $\mathcal B_{\max}$:
\begin{equation}
\mathcal{B}_{\max} = 2 \sqrt{m_1 + m_2},
\end{equation}
where $m_1$ and $m_2$ are the two largest eigenvalues of $M$.

If the density matrix $\rho$ is completely reconstructed, the
concurrence and the CHSH parameter can be directly computed from it.
However, in the present channel the missing neutrinos induce the null
direction $C_{nr}^-=C_{nr}-C_{rn}$, so this component cannot be
determined from the visible pion distribution alone. At first sight,
this seems to prevent a unique determination of nonlinear quantum
observables.

Nevertheless, the density matrix is not arbitrary. It has to satisfy
the physical positivity condition
\(\rho\succeq 0,\)
or equivalently, all its eigenvalues must be non-negative.
We therefore treat the null component $C_{nr}^-$ as an unconstrained
variable, while fixing the remaining $14$ spin
parameters to the values obtained from the convergent iteration. The
positivity condition then determines an allowed interval for
$C_{nr}^-$ in each production-angle bin. Scanning this interval further
gives the corresponding physical ranges of the concurrence and the
CHSH parameter.

In practice, finite-sample statistical fluctuations can make the reconstructed central values slightly incompatible with exact positivity when all identifiable components are held fixed. We therefore introduce a small numerical tolerance and require
\[
r_k\geq -\varepsilon,\qquad k=1,2,3,4,
\]
where $r_k$ are the eigenvalues of the density matrix and $\varepsilon$ is a small positive number. A smaller $\varepsilon$ gives tighter intervals, whereas an excessively small value may leave no feasible interval because of finite-sample fluctuations. In the numerical results, we take $\varepsilon=10^{-3}$, which provides a balance between interval tightness and numerical feasibility. This tolerance is introduced solely to accommodate finite-sample statistical fluctuations and should not be interpreted as a physical relaxation of density-matrix positivity.

In addition to the positivity-constrained intervals, Fig.~\ref{fig:tau-quantum-observables} shows illustrative point estimates obtained by directly evaluating the concurrence and the CHSH parameter from the iterated density matrix.
In this direct evaluation, the iterated value
of the null component $C_{nr}^-$ is used together with the identifiable
components. Strictly speaking, this central-value prescription is not
part of the model-independent reconstruction, because $C_{nr}^-$ is not
determined by the visible distribution. 
In the SM closure test, however, the iteration is initialized with all spin-correlation components set to zero, while the SM truth value of the null component is also zero.
As a result, the iterated $C_{nr}^-$ remains close to
zero, up to statistical fluctuations. Once the identifiable components
have converged, the directly computed central values therefore approach
the truth values and lie inside the positivity-constrained physical
ranges.

These point estimates may be viewed as illustrative results obtained after imposing the additional assumption that the null component remains near its SM value, $C_{nr}^-=0$. Without this assumption, only the positivity-constrained ranges represent the model-independent information contained in the visible pion data.

\bibliography{mainrefs}
\bibliographystyle{JHEP}

\end{document}